\documentclass[11pt,reqno]{amsart}
\usepackage{amsfonts}
\usepackage[framemethod=TikZ]{mdframed}
\usepackage{xcolor}
\usepackage{framed}
\usepackage{geometry}
\usepackage{amsmath,amssymb,amsthm}
\usepackage[authoryear]{natbib}
\usepackage{graphicx}
\usepackage{comment}
\usepackage{tikz}
\usepackage{bm}
\usepackage{epstopdf}
\usepackage{hyperref}

\theoremstyle{plain}

\theoremstyle{definition}

\newtheorem{remark}{Remark}
\newtheorem{example}{Example}
\newtheorem{algorithm}{Algorithm}

\renewcommand{\Pr}{{\mathrm{P}}}

\numberwithin{equation}{section}
 
\usetikzlibrary {positioning}

\def\argmin{\mathop{\arg\min}}

\input{tcilatex}

\begin{document}
\title[Extremal Quantile Regression]{Extremal Quantile Regression: An Overview}
\thanks{MIT, BU, and MIT. We would like to thank Roger Koenker for useful comments. We would like to
acknowledge the financial support from the NSF}
\author{Victor Chernozhukov \ Iv\'an Fern\'andez-Val \ Tetsuya Kaji}

\date{\today }

\begin{abstract}
Extremal quantile regression, i.e. quantile regression applied to the tails of the conditional distribution, counts with an increasing number of economic and financial applications such as value-at-risk, production frontiers, determinants of low infant birth weights, and auction models. This chapter provides an overview of recent developments in the theory and empirics of extremal quantile regression. The advances in the theory have relied on the use of extreme value approximations to the law of the Koenker and Bassett (1978) quantile regression estimator.  Extreme value laws not only have been shown to provide more accurate approximations than Gaussian laws at the tails, but also have served as the basis to develop bias corrected estimators and inference methods using simulation and suitable variations of bootstrap and subsampling.  The applicability of these methods is illustrated with two empirical examples on conditional value-at-risk and financial contagion.
\end{abstract}

\maketitle

\section{Introduction}

In 1895, the Italian econometrician Vilfredo Pareto discovered that  the power law describes well the tails of income and wealth data. This simple observation stimulated further applications of the power law to economic data including \citet{zipf:1949}, \citet{mandelbrot:1963}, \citet{fama:1965}, \citet{praetz:1972}, \citet{sen:1973}, and \citet{longin:1996}, among many others. It also opened up a theory to analyze the properties of the tails of the distributions so-called Extreme Value (EV) theory, which  was developed by \citet{gnedenko:1943} and \citet{dehaan:1970}.  \citet{jansen:1991} applied this theory to analyze the tail properties of US financial returns and concluded that the 1987 market crash was not an outlier; rather, it was a rare event whose magnitude could have been predicted by prior data. This work stimulated numerous other studies that rigorously documented the tail properties of economic data \citep{embrechts:1997}. 

\medskip

\citet{victor:annals} extended the EV theory to develop extreme quantile regression models in the tails, and analyze the properties of the  \citet{koenker:1978} quantile regression estimator,  called \textit{extremal quantile regression}. This work builds especially upon  \citet{feigin:1994} and  \citet{knight:linear}, which studied the most extreme, frontier regression case in the location model.  Related results for the frontier case  -- the regression frontier estimators -- were developed by  \citet{smith:1994}, \citet{chernozhukov:1998}, \citet{jureckova:1999},  and \citet{portnoy:jur}.   \citet{portnoy-koenker:1989} and \citet{gjkp:1993} implicitly contained some results on extending the normal approximations to intermediate order regression quantiles (moderately extreme quantiles) in location models. The complete theory for intermediate order regression quantiles was developed in \citet{victor:annals}. \citet{jureckova:2016} recently characterized properties of averaged extreme regression quantiles.

%

\medskip

In this chapter we review the theory of extremal quantile regression.
  We start by introducing the general setup that will be used throughout the chapter. Let $Y$ be a continuous response variable of interest with 
distribution function $F_Y(y) = \Pr(Y \leq y)$. The marginal $\tau$-quantile of
$Y$  is the left-inverse of $y \mapsto F_Y(y)$ at $\tau$, that is $Q_Y(\tau) := \inf\{y: F_{Y}(y) \geq \tau\}$ for some $\tau \in (0,1).$ Let $X$ be a $d_x$-dimensional vector of covariates related to $Y$, $F_X$ be the distribution function of $X$, and
$F_{Y}(y | x)= \Pr(Y \leq y | X=x)$ be the conditional distribution
function of $Y$ given $X=x$. The conditional $\tau$-quantile of $Y$
given $X=x$ is the left-inverse of $y \mapsto F_{Y}(y | x)$ at $\tau$, that is $Q_Y(\tau | x) := \inf\{y: F_{Y}(y | x) \geq \tau\}$ for some $\tau \in (0,1).$  We refer to $x \mapsto Q_{Y}(\tau|x)$ as the
$\tau$-quantile regression function.  This function measures the
effect of $X$ on $Y$, both at the center and at the tails of the outcome distribution. A marginal or conditional
$\tau$-quantile is \textit{extremal} whenever the probability index
$\tau$ is either close to zero or close to one. Without loss of generality, we focus the discussion
on $\tau$ close to zero.

\medskip

The analysis of the properties of the estimators of extremal quantiles relies on EV theory. This theory uses sequences of quantile indexes $\{\tau_T\}_{T=1}^{\infty}$ that change with the sample size $T$. Let  $\tau_T T$
be the order of the $\tau_T$-quantile. A sequence of quantile index and sample size pairs
$\{\tau_T, T\}_{T=1}^\infty$ is said to be an {\it extreme order}
sequence if $\tau_T
 \searrow 0$ and $ \tau_T T \rightarrow k \in (0, \infty)$ as $T \to \infty$;
 an  {\it intermediate order} sequence if $\tau_T
 \searrow 0$ and $ \tau_T T \rightarrow \infty$ as $T \to \infty$;  and a {\it central order} sequence if $\tau_T$
 is fixed as $T \rightarrow \infty$.  In this chapter we show that each of these sequences produce different asymptotic approximations to the distribution of the quantile regression estimators.  The extreme order
 sequence leads to an EV law in large samples, whereas the intermediate and central
 sequences lead to normal laws. The EV law provides a better approximation to the extremal quantile regression estimators.

\medskip

We conclude this introductory section with a review of some applications of extremal quantile regression to economics and finance.
\begin{example}[Conditional Value-at-Risk]
The Value-at-Risk (VaR) analysis seeks to forecast or explain low quantiles of future portfolio returns of an institution, $Y$,  using  current information, $X$ \citep{vl,engle:2004}.  Typically, the extremal $\tau$-quantile regression functions $x \mapsto Q_Y(\tau | x)$ with $\tau=0.01$ and $\tau=0.05$ are of interest. 
The VaR is a risk measure commonly used in real-life financial management, insurance, and actuarial science \citep{embrechts:1997}.
We provide an empirical example of VaR in Section \ref{sec:empirics}.
\end{example}

\begin{example}[Determinants of Birthweights] In health economics, we may be interested in how smoking, absence of prenatal care, and other maternal behavior during pregnancy, $X$, affect infant birthweights, $Y$ \citep{abrevaya}.
Very low birthweights are connected with subsequent health problems and therefore extremal quantile regression can help identify factors to improve adult health outcomes. 
\citet{chernozhukov:2011} provide an empirical study of the determinants of extreme birthweights.
\end{example}

\begin{example}[Probabilistic Production Frontiers]
An important application to industrial organization is the determination of efficiency or production frontiers, pioneered by \citet{aigner:1968}, \citet{timmer}, and \citet{aigner:1976}. Given the cost of production and possibly other factors, $X$, we are interested in the highest production levels, $Y$, that only a small fraction of firms, the most efficient firms, can attain. These (nearly) efficient production levels can be formally described by the extremal $\tau$-quantile regression function $x \mapsto Q_Y(\tau | x)$ for $\tau \in [1-\varepsilon, 1)$ and $\varepsilon > 0$; so that only a $\varepsilon$-fraction of firms produce $Q_Y(\tau | X)$ or more.  
\end{example}

\begin{example}[Approximate Reservation Rules] In labor economics, \citet{flinn:1982} proposed a  job search  model with approximate reservation rules. The reservation rule measures the wage level, $Y$, below which a  worker with characteristics, $X$,  accepts a job with small probability $\varepsilon$, and can by described by the extremal $\tau$-quantile regression $x \mapsto Q_Y(\tau | x)$ for $\tau \in (0,\varepsilon]$.
\end{example}

\begin{example}[Approximate $(S,s)$-Rules] The $(S,s)$-adjustment models arise as an optimal policy in many economic models \citep{arrow:ss}. For example, the capital stock, $Y$, of a firm with characteristics $X$ is adjusted sharply up to the level $S(X)$ once it has depreciated below some low level $s(X)$ with  probability close to one, $1-\varepsilon$. This conditional $(S,s)$-rule can be described by the extremal $\tau$-quantile regression functions $x \mapsto Q_Y(\tau|x)$ and $x \mapsto Q_Y(1-\tau|x)$ for $\tau \in (0,\varepsilon]$.

\end{example}

\begin{example}[Structural Auction Models] Consider a first-price procurement auction where bidders hold independent valuations.  \citet{dp:superconsistent} modelled the winning bid, $Y$, as $Y = c(Z) \beta(N) + \varepsilon$,  where $c(Z)$ is the efficient cost function that depends on the bid characteristics, $Z$, $\beta(N) \geq 1$ is a mark-up that approaches 1 as the number of bidders $N$ approaches infinity, and the disturbance $\varepsilon$ captures small bidding mistakes independent of $Z$ and $N$. By construction, the structural function $(z,n) \mapsto c(z) \beta(n)$ corresponds to the extremal quantile regression function $x \mapsto Q_Y(\tau | x)$ for $x = (z,n)$ and $\tau  \in (0 , \Pr(\varepsilon \leq 0)]$. 
\end{example}

\begin{example}[Other Recent Applications] Following the pioneering work of \citet{powell:1984}, \citet{altonji:2012}  applied extremal quantile regression to estimate extensive margins of demand functions with corner solutions.  \citet{d'haultfoeuille:2015} used extremal quantile regression to deal with endogenous sample selection under the assumption that there is no selection at very high quantiles of the response variable $Y$ conditional on covariates $X$. They applied this approach to the estimation of  the black wage gap for  young males in the US. \citet{zhang:2015} employed extremal quantile regression methods to estimate tail quantile treatment effects  under a selection on observables assumption.

\end{example}


\medskip

\textbf{Notation:} The symbol $\to_d$ denotes convergence in law. For two real numbers $a,b$, $a \ll b$ means that $a$ is much less than $b$. More notation will be introduced when it is first used.

\medskip

\textbf{Outline:} The rest of the chapter is organized as follows. Section \ref{sec:model} reviews models for marginal and conditional extreme  quantiles. Section  \ref{sec:inference} describes estimation and inference methods for extreme quantile models. Section \ref{sec:empirics} presents two empirical applications of extremal quantile regression to conditional value-at-risk and financial contagion.

\medskip

\section{Extreme Quantile Models}\label{sec:model}
This section reviews  typical modeling assumptions in extremal quantile regression. They embody Pareto conditions on the tails of the distribution of the response variable $Y$ and linear specifications for the $\tau$-quantile regression function $x \mapsto Q_Y(\tau|x)$.

\subsection{Pareto-Type and Regularly Varying Tails} \label{subsection:tails}
The theory for extremal quantiles often assumes that the tails of the distribution of $Y$ have Pareto-type behavior, meaning that the tails decay approximately as a power function, or more formally, a regularly varying function.
The Pareto-type tails encompass a rich variety of tail behaviors, from thick to thin tailed distributions, and from bounded to unbounded support distributions.

\medskip
:
Define the variable $U$ by $U := Y$ if the lower end-point of the support of $Y$ is $-\infty$ and by $U := Y-Q_Y(0)$ if the lower end-point of the support of $Y$ is finite.
In words, $U$ is a shifted copy of $Y$ whose support ends at either $-\infty$ or $0$.
The assumption that the random variable $U$ exhibits a {\em Pareto-type tail} is stated by the following two equivalent conditions:%
\footnote{$a \sim b$ means that $a/b \to 1$ with an appropriate notion of limit.}
\begin{alignat}{4}
	Q_U(\tau) &\sim L(\tau) \cdot \tau^{-\xi} &\qquad \text{as} \qquad && \tau &\searrow 0, \label{power1} \\
	F_U(u) &\sim \bar{L}(u) \cdot u^{-1/\xi} &\qquad \text{as} \qquad && u &\searrow Q_U(0), \label{power2}
\end{alignat}
for some $\xi \neq 0$, where $\tau \mapsto L(\tau)$ is a non-parametric slowly-varying function at $0$, and $u \mapsto \bar{L}(u)$ is a non-parametric slowly-varying function at $Q_U(0)$.%
\footnote{A function $z \mapsto f(z)$ is said to be {\em slowly-varying} at $z_0$ if $\lim_{z \searrow z_0} f(z)/f(mz) = 1$ for every $m > 0$}
The leading examples of slowly-varying functions are the constant function and the logarithmic function.
The number $\xi$ as defined in (\ref{power1}) or (\ref{power2}) is called the {\em extreme value (EV) index} or the {\em tail index}.

\medskip

The absolute value of $\xi$ measures the heavy-tailedness of the distribution.
The support of a Pareto-type tailed distribution necessarily has a finite lower bound if $\xi < 0$ and an infinite lower bound if $\xi > 0$.
Distributions with $\xi>0$ include stable, Pareto, Student's $t$, and many others. For example, the $t$-distribution with $\nu$ degrees of freedom has $\xi=1/\nu$ and exhibits a wide range of tail behaviors.
In particular, setting $\nu =1$ yields the Cauchy distribution which has heavy tails with $\xi=1$, while setting $\nu =30$ gives a distribution that has light tails with $\xi =1/30$ and is very close to the normal distribution.  On the other hand, distributions with $\xi < 0$ include the uniform, exponential, Weibull, and many others.

\medskip

The assumption of Pareto-type tails can be equivalently cast in terms of a regular variation assumption, as is commonly done in the EV theory.  A distribution function $u \mapsto F_U(u)$ is said to be {\em regularly varying} at $u = Q_U(0)$ with index of regular variation $-1/\xi$ if $$\lim_{y\searrow Q_U(0)} F_U(ym) / F_U(y) = {m}^{-1/\xi} \qquad \text{for every} \qquad m > 0.$$  This condition is equivalent to the regular variation of the quantile function $\tau \mapsto Q_U(\tau)$ at $\tau = 0$ with index $-\xi$, $$\lim_{\tau \searrow 0} Q_U(\tau m) / Q_U(\tau) = {m}^{-\xi} \qquad \text{for every} \qquad m > 0.$$

\medskip

The case of $\xi=0$ corresponds to the class of rapidly varying distribution functions.  Such distribution functions have exponentially light tails, with the normal and exponential distributions being the chief examples.
For the sake of simplicity, we omit this case from our discussion.
Note, however, that since the limit distribution of the main statistics is continuous in $\xi$, the inference theory for $\xi = 0$ is included by taking $\xi \to 0$.


\medskip

%
%

\subsection{Extremal  Quantile Regression Models}

The most common model for the quantile regression (QR) function  is the linear in parameters specification
\begin{equation} \label{eq1}
		Q_Y(\tau | x) = B(x)' \beta(\tau) \qquad \text{for all $\tau \in (0,\eta]$ and some $\eta \in (0,1)$},
\end{equation}
and for every $x \in \mathbf{X}$, the support of $X$.
This linear functional form not only provides computational convenience but also has good approximation properties. Thus,  the set  $B(x)$ can include transformations of $x$ such as polynomials, splines, indicators or interactions such that $x \mapsto B(x)' \beta(\tau)$ is close to $x \mapsto Q_Y(\tau | x)$. 
In what follows,   without loss of generality we lighten the notation by using $x$ instead of $B(x)$. We also assume that the $d_x$-dimensional vector $x$ contains a constant as the first element, has a compact support $\mathbf{X}$, and satisfies the regularity conditions stated in Assumption 3 of \citet{chernozhukov:2011}.
Compactness is needed to ensure the continuity and robustness of the mapping from extreme events in $Y$ to the extremal QR statistics. Even if $\mathbf{X}$ is not compact, we can select the data for which $X$ belongs to a compact region.

\medskip

The main additional assumption for extremal quantile regression is that $Y$, transformed by some auxiliary regression line $X'\beta_e$, has Pareto-type tails.
More precisely, together with (\ref{eq1}), it assumes that there exists an auxiliary regression parameter $\beta_e \in \mathbb{R}^{d_x}$ such that the disturbance $V := Y - X'\beta_e$ has lower end point $s = 0$ or $s = -\infty$ a.s., and its conditional quantile function $Q_V(\tau | x)$ satisfies the tail equivalence relationship:
\begin{equation} \label{location-scale}
	Q_V(\tau | x) \sim x' \gamma \cdot Q_U(\tau) \qquad \text{as $\tau \searrow 0$ uniformly in $x \in \mathbf{X} \subseteq \mathbb{R}^{d_x}$},
\end{equation}
for some quantile function $Q_U(\tau)$ that exhibits a Pareto-type tail (\ref{power1}) with EV index $\xi$, and some vector parameter $\gamma$ such that $E[X]' \gamma = 1$ and $X' \gamma > 0$ a.s.

\medskip

Condition (\ref{location-scale}) imposes a location-scale shift model. This model is more general than the standard location shift model that replaces $x' \gamma$ by a constant, because it permits conditional heteroskedasticity that is common in economic applications. Moreover,  condition (\ref{location-scale})  only affects the far tails, and therefore allows covariates to affect extremal and central quantiles very differently.
Even at the tails, the local effect of the covariates is approximately given by $\beta(\tau) \approx \beta_e + \gamma Q_U(\tau)$, which can be heterogenous across extremal quantiles.

\medskip

Existence and Pareto-type behavior of the conditional quantile density function is also often imposed as a technical assumption that facilitates the derivation of inference results. Accordingly, we will assume that the conditional quantile density function $\partial Q_V(\tau | x) / \partial \tau$ exists and satisfies the tail equivalence relationship 
$$\partial Q_V(\tau | x) / \partial \tau \sim x' \gamma \cdot \partial Q_U(\tau) / \partial \tau \qquad \text{as} \qquad  \tau \searrow 0$$ uniformly in $x \in \mathbf{X},$ where $\partial Q_U(\tau) / \partial \tau$ exhibits Pareto-type tails as $\tau \searrow 0$ with EV index $\xi+1$.

\medskip

\section{Estimation and Inference Methods}\label{sec:inference} This section reviews estimation and inference methods for extremal quantile regression. Estimation is based on the \citet{koenker:1978} quantile regression estimator. We consider both analytical and resampling methods. These methods are introduced  in the univariate case of marginal quantiles and then extended to the multivariate or regression case of conditional quantiles. We start by imposing some general sampling conditions.

\subsection{Sampling Conditions} We assume that we have a sample of $(Y,X)$ of size $T$ that is either independent and identically distributed (i.i.d.) or stationary and weakly-dependent, with extreme events satisfying a non-clustering condition. In particular, the sequence $\{(Y_t, X_t) \}_{t=1}^T$ is assumed to form a stationary, strongly mixing process with geometric mixing rate, that satisfies the condition that curbs clustering of extreme events \citep{chernozhukov:2011}. The assumption of mixing dependence is standard in econometrics \citep{white:2001}.
The non-clustering condition is of the \citet{meyer:1973} type and states that the probability of two extreme events co-occurring at nearby dates is much lower than the probability of just one extreme event.
For example, it assumes that a large market crash is not likely to be immediately followed by another large crash.
This assumption is convenient because it leads to limit distributions of extremal quantile regression estimators as if independent sampling had taken place. The plausibility of the non-clustering assumption is an empirical matter.
 
\medskip

\subsection{Univariave Case: Marginal Quantiles}\label{sec:univariate}


%
%
%

The analog estimator of the marginal $\tau$-quantile is the sample $\tau$-quantile,
$$
\hat{Q}_Y(\tau) = Y_{(\lfloor \tau T \rfloor)},
$$
where $Y_{(s)}$ is the $s^{th}$ order statistic of $(Y_1, \ldots, Y_T)$, and $\lfloor z \rfloor$ denotes the integer part of $z$. The sample $\tau$-quantile can also be computed as a solution to an optimization program,
\begin{equation*} \label{argmin}
	\hat Q_Y(\tau) \in \argmin_{\beta \in \mathbb{R}} \, \sum_{t=1}^T \rho_\tau(Y_t - \beta),
\end{equation*}
where $\rho_\tau(u) := (\tau - 1\{u < 0\}) u$ is the asymmetric absolute deviation function of \citet{fox:rubin}.

\medskip

We review the asymptotic behavior of the sample quantiles under extreme and intermediate order sequences, and describe inference methods for extremal marginal quantiles.

\medskip

\subsubsection{Extreme Order Approximation} \label{univariate:extreme} Recall the following classical result from EV theory on the limit distribution of extremal order statistics: 
as $T \to \infty$, for any integer $k\geq 1$ such that $k_T := \tau_T T \to k$,
\begin{equation} \label{gnedenko}
	\hat{Z}_T(k_T) = A_T (\hat Q_Y(\tau_T)-Q_Y(\tau_T)) \ \to_d \ \hat{Z}_\infty(k) = \Gamma_k^{-\xi} - k^{-\xi},
\end{equation}
where
\begin{equation}\label{eq:cnq}
	A_T = 1/ Q_U(1/T), \qquad \Gamma_k = \mathcal{E}_1 + \cdots + \mathcal{E}_k.
\end{equation}
The variable $U$ is defined in Section \ref{subsection:tails}
and $\{\mathcal{E}_1, \mathcal{E}_2, \cdots\}$ is an i.i.d.\ sequence of standard exponential variables.
We call $\hat{Z}_T(k_T)$ the {\em canonically-normalized quantile (CN-Q) statistic} because it depends on the canonical scaling constant $A_T$. The result (\ref{gnedenko}) was obtained by \citet{gnedenko:1943} for  i.i.d. sequences of random variables. It continues to hold for stationary weakly-dependent series, provided the probability of extreme events occurring in clusters is negligible relative to the probability of a single extreme event \citep{meyer:1973}. \citet{leadbetter}  extended the result to more general time series processes. 

\medskip

The limit in (\ref{gnedenko}) provides an EV distribution as an approximation to the finite-sample distribution of $\hat Q_Y(\tau_T)$, given the scaling constant $A_T$. The limit distribution is characterized by the  EV index $\xi$  and the variables $\Gamma_k$. The index $\xi$ is usually unknown but can be estimated by one of the methods described below.   The variables $\Gamma_k$ are {\it gamma} random variables. 
The limit variable $\hat{Z}_\infty(k)$ is therefore a transformation of gamma variables, which has finite mean if $\xi < 1$ and has finite moments of up to order $1 / \xi$ if $\xi > 0$. Moreover, the limit EV distribution  is not symmetric, predicting significant median asymptotic bias in $\hat Q_Y(\tau_T)$ with respect to $Q_Y(\tau_T)$ that  motivates the use of the median-bias correction techniques discussed below.

\medskip

The classical result (\ref{gnedenko}) is often  not feasible for inference on $Q_Y(\tau_T)$ because  the constant $A_T$ is unknown and generally cannot be estimated consistently \citep{bertail:2004}.
One way to deal with this problem is to add strong parametric assumptions on the non-parametric, slowly varying function $\tau \mapsto L(\tau)$ in Equation (\ref{power1}) in order to estimate $A_T$ consistently.
This approach is discussed in Section \ref{univariate:ev index}. An alternative is to consider the  {\em self-normalized quantile (SN-Q) statistic}
\begin{equation} \label{SN}
Z_T(k_T) := \mathcal{A}_T (\hat{Q}_Y(\tau_T) - Q_Y(\tau_T)), \ \ \mathcal{A}_T := \frac{\sqrt{k_T}}{\hat{Q}_Y(m\tau_T) - \hat{Q}_Y(\tau_T)},
\end{equation}
for $m> 1$ such that $mk$ is an integer. For example, $m = p/k_T + 1 = p/k + 1 + o(1)$ for some spacing parameter $p \geq 1$ (e.g.\ $p = 5$). The scaling factor $\mathcal{A}_T$ is completely a function of data and therefore feasible, avoiding the need for the consistent estimation of $A_T$.  \citet{chernozhukov:2011} shows that as $T \to \infty$, for any integer $k\geq 1$ such that $k_T := \tau_T T \to k$, 
\begin{equation} \label{EV approximation}
	Z_T(k_T)  \ \to_d \ Z_\infty(k) := \frac{\sqrt{k} (\Gamma_k^{-\xi} - k^{-\xi})}{\Gamma_{mk}^{-\xi} - \Gamma_k^{-\xi}},
\end{equation}
The limit distribution in (\ref{EV approximation}) only depends on the EV index $\xi$, and its quantiles can be easily obtained by the resampling methods described in Section \ref{univariate:cv}.

\medskip

\subsubsection{Intermediate Order Approximation} \label{univariate:intermediate}
\citet{Dekkers:dehaan} show that as $\tau \searrow 0$ and $k_T \to \infty$, and under further regularity conditions,
\begin{equation} \label{normal approximation}
	Z_T(k_T) := \mathcal{A}_T (\hat{Q}_Y(\tau_T) - Q_Y(\tau_T)) \ \to_d \ \mathcal{N} \biggl( 0, \, \frac{\xi^2}{(m^\xi - 1)^2} \biggr),
\end{equation}
where $\mathcal{A}_T$ is defined as in (\ref{SN}).
This result yields a normal approximation to the finite-sample distribution of $\hat{Q}_Y(\tau)$. 
Note this normal approximation holds only when $k_T \to \infty$, while the EV approximation (\ref{EV approximation}) holds not only when $k_T \to k$ but also when  $k_T \to \infty$ because the EV distribution converges to the normal distribution as $k \to \infty$.
In finite samples, we may interpret the condition $k_T \to \infty$ as requiring that $k_T \geq 30$.

\medskip

\begin{figure}[!http!]
\includegraphics[width=\textwidth]{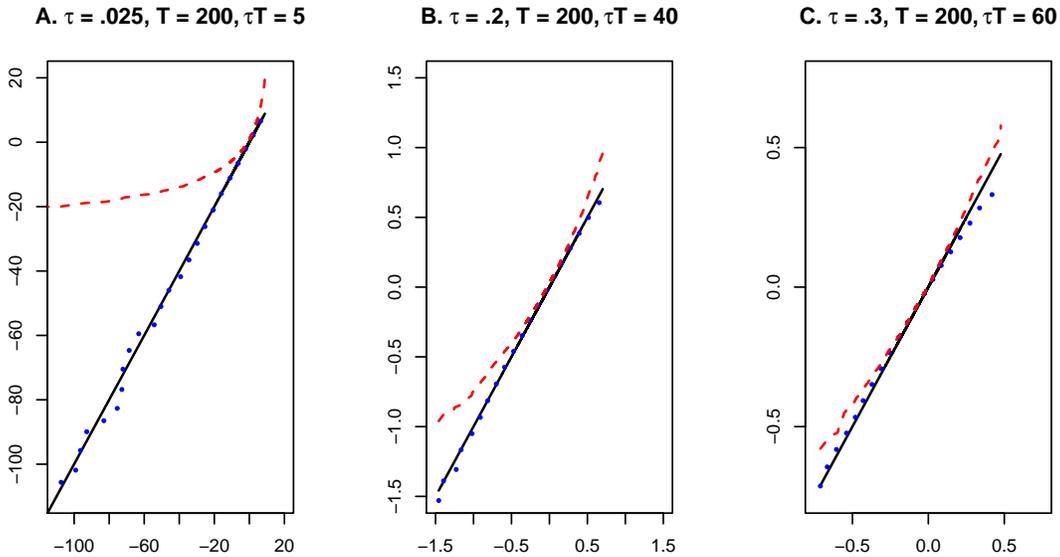}
 \caption{Quantiles of the distribution of the sample quantiles  vs. quantiles of EV and normal
 approximations. The figure is based on a design where $Y$
 follows a Cauchy distribution.  The solid line ``------" shows the quantiles of the exact
distribution of the sample $\tau$-quantile with $\tau \in \{ .025, .2, .3\},$
obtained from 10,000 simulations.  The dashed line ``- \ - \ -" shows the
quantiles of the normal approximation. The dotted line
``......" shows the quantiles of EV approximation.}\label{shock}
\end{figure}

Figure \ref{shock}, taken from   \citet{chernozhukov:2011}, shows that the EV distribution provides a better approximation to the distribution of  extremal  sample quantiles than the normal distribution when $k_T < 30$. It plots the quantiles of these distributions against the quantiles of the finite-sample distribution of the sample $\tau$-quantile for  $T=200$ and $\tau \in \{ .025, .2, .3\}$. If either
the EV  or the normal distributions were to coincide with the exact distribution,
then their quantiles would fall on the 45 degree line shown
by the solid line. When the order $\tau T$ is
$5$ or $40$, the quantiles of the EV distribution are very close to the 45
degree line, and in fact are much closer to this line than the
quantiles of the normal distribution. Only for the case when the 
order $\tau T$ becomes $60$, do the quantiles of the EV and
normal distributions become comparably close to the 45 degree line.

\medskip

\subsubsection{Estimation of $\xi$} \label{univariate:ev index}

Some inference methods for extremal marginal quantiles require a consistent estimator of the EV index $\xi$.
We describe two well-known estimators of $\xi$. The first estimator, due to \citet{pickands:1975}, relies on the ratio of sample quantile spacings:
\begin{equation}\label{eq:pickands}
	\hat{\xi}_P := \frac{-1}{\ln 2} \ln \biggl( \frac{\hat{Q}_Y(4\tilde{\tau}_T) - \hat{Q}_Y(2\tilde{\tau}_T)}{\hat{Q}_Y(2\tilde{\tau}_T) - \hat{Q}_Y(\tilde{\tau}_T)} \biggr).
\end{equation}
Under further regularity conditions, for $\tilde{\tau}_T \searrow 0$ and $\tilde{\tau}_T T \to \infty$ as $T \to \infty$,
\[
	\sqrt{\tilde{\tau}_T T} (\hat{\xi}_P - \xi) \ \to_d \ \mathcal{N} \biggl( 0, \, \frac{\xi^2(2^{2\xi+1}+1)}{[2 (2^{\xi}-1) \ln 2]^2} \biggr)
\]
The second estimator, developed by \citet{hill}, is the moment estimator:%
\begin{equation}\label{eq:hill}
	\hat{\xi}_H := -\frac{\sum_{t=1}^T 1\{Y_t < \hat{Q}_Y(\tilde{\tau}_T)\} \ln(Y_t/\hat{Q}_Y(\tilde{\tau}_T))}{\tilde{\tau}_T \sum_{t=1}^T 1\{Y_t < \hat{Q}_Y(\tilde{\tau}_T)\}},
\end{equation}
which is applicable when $\xi>0$ and $\hat{Q}_Y(\tilde{\tau}_T) < 0$.
This estimator is motivated by the maximum likelihood method that fits an exact power law to the tail data.
Under further regularity conditions,  for $\tilde{\tau}_T \searrow 0$ and $\tilde{\tau}_T T \to \infty$ as $T \to \infty$,
\[
	\sqrt{\tilde{\tau}_T T} (\hat{\xi}_H - \xi) \ \to_d \ \mathcal{N}(0, \xi^2).
\]
The previous limit results can be used to construct the confidence intervals and median-bias corrections for $\xi$. We give an example of these confidence intervals and corrections in Section \ref{sec:empirics}.

\medskip

\citet{embrechts:1997} provide methods for choosing $\tilde{\tau}_T$. There is a variance-bias  trade-off, as the variance of the estimator decreases,  but the bias increases, as $\tilde{\tau}_T$ increases.
Another view on the choice of $\tilde{\tau}_T$ is that the statistical models are approximations, but not literal descriptions of the data.
In practice, the dependence of $\hat{\xi}$ on the threshold $\tilde{\tau}_T$ reflects that power laws with different values of $\xi$ would fit some tail regions better than others.
Therefore, if the interest lies in making the inference on $Q_Y(\tau_T)$ for a particular $\tau_T$, it seems reasonable to use $\hat{\xi}$ constructed using  $\tilde{\tau}_T = \tau_T$ or the closest $\tilde{\tau}_T$ to $\tau_T$ subject to  $\tilde{\tau}_T T \geq 30$. This condition ensures to have a sufficient number of observations to estimate $\xi$.

\medskip

\subsubsection{Estimation of $A_T$} To use the CN-Q statistic for inference, we need to estimate the scaling constant $A_T$ defined in \eqref{eq:cnq}.
This requires additional strong restrictions on the underlying model.
For instance, assume that the slowly varying function $\tau \mapsto L(\tau)$ is just a constant $L$, i.e., as $\tau \searrow 0$,
\begin{equation} \label{univariate:param rest}
	1/Q_U(\tau) = L \cdot \tau^{\xi} \cdot (1 + \delta(\tau)) \quad \text{for some $L \in \mathbb{R}$ and $\delta(\tau) \to 0$.}
\end{equation}
Then, we can estimate $L$ by
\begin{equation} \label{univariate:L estimate}
	\hat{L} :=  \frac{\hat{Q}_Y(2\tilde{\tau}_T) - \hat{Q}_Y(\tilde{\tau}_T)}{(2^{-\hat{\xi}}-1) \cdot \tilde{\tau}_T^{-\hat{\xi}}},
\end{equation}
where $\hat{\xi}$ is either the Pickands or Hill estimator given in \eqref{eq:pickands} and \eqref{eq:hill}, and $\tilde{\tau}_T$ can be chosen using the same methods as in the estimation of $\xi$.
Then, the estimator of $A_T$ is 
\begin{equation} \label{univariate:A estimator}
	\hat{A}_T := \hat{L} \cdot T^{-\hat{\xi}}.
\end{equation}

\medskip

\subsubsection{Computing Quantiles of the Limit EV Distributions}\label{univariate:cv}

The inference and bias corrections for extremal quantiles are based on the  EV approximations given in \eqref{gnedenko} and \eqref{EV approximation}, with an estimator in place of the EV index if needed.
In practice, it is convenient to compute the quantiles of the EV distributions using simulation or resampling methods, instead of an analytical method. Here we illustrate two of such methods: extremal bootstrap and extremal subsampling. The bootstrap method relies on simulation, whereas the subsampling method relies on drawing subsamples from the original sample. Subsampling has the advantages that it does not require the estimation of $\xi$ and is consistent under general conditions (e.g.\ subsampling does not require i.i.d.\ data).  Nevertheless, bootstrap is more accurate than subsampling when a stronger set of assumptions holds. It should be noted here that the empirical or nonparametric bootstrap is not consistent for extremal quantiles \citep{bickel}.

\medskip

The extremal bootstrap is based on simulating samples from a random variable with the same tail behavior as $Y$. Consider the random variable,\footnote{The variable $Y^*$ follows the generalized extreme value distribution, which nests the Frechet, Weibull, and Gumbell distributions. There are other possibilities, for example, the standard exponential distribution can be replaced by the standard uniform distribution, in which case $Y$ would follow the generalized Pareto distribution.}
\begin{equation} \label{generalized pareto}
	Y^* = \frac{\mathcal{E}^{-\xi}-1}{-\xi},  \qquad \mathcal{E} \sim \text{Exponential}(1).
\end{equation}
This variable has the quantile function 
\begin{equation} \label{quant:gp}
	Q_{Y^*}(\tau) =  \frac{[-\ln (1-\tau)]^{-\xi} -1}{-\xi},
\end{equation}
which satisfies Condition (\ref{power2})  because $Q_{Y^*}(\tau) - 1/\xi \sim \tau^{-\xi}/\xi$.
The extremal bootstrap  estimates the distribution of $Z_T(k_T) = \mathcal{A}_T (\hat{Q}_Y(\tau_T) - Q_Y(\tau_T))$ by the  distribution of $Z^*_T(k_T) =  \mathcal{A}_T (\hat{Q}_{Y^*}(\tau_T) - Q_{Y^*}(\tau_T))$ obtained by simulation. This approximation reproduces both the EV limit (\ref{EV approximation}) under extreme order sequences and the normal limit (\ref{normal approximation}) under  intermediate order sequences. 
Algorithm \ref{alg:eb} describes the implementation of this method.

\begin{algorithm}[Extremal Bootstrap]\label{alg:eb} (1) Choose the quantile index of interest $\tau_T$ and the number of simulations $S$ (e.g., $S=500$). (2) For each $s \in \{1, \ldots, S\}$, draw an i.i.d. sequence $\{Y^*_{1,s}, \dots, Y^*_{T,s}\}$ from the random variable $Y^*$ defined in  \eqref{generalized pareto}, replacing $\xi$ by the estimator  \eqref{eq:pickands}  or \eqref{eq:hill}. (3) For each $s \in \{1, \ldots, S\}$, compute the statistic $Z^*_{T,s}(k_T) = \mathcal{A}_T (\hat{Q}_{Y^*}(\tau_T) - Q_{Y^*}(\tau_T))$, where $\hat{Q}_{Y^*}(\tau_T)$ is the sample $\tau_T$-quantile in the bootstrap sample $\{Y^*_{1,s}, \dots, Y^*_{T,s}\}$ and $ Q_{Y^*}(\tau_T)$ is defined as in \eqref{quant:gp} replacing $\xi$ by the same estimator as in step (2). (4) Estimate the quantiles of $Z_T(k_T) = \mathcal{A}_T (\hat{Q}_Y(\tau_T) - Q_Y(\tau_T))$ by the sample quantiles of $\{Z^*_{T,s}(k_T)\}_{s=1}^S$.
\end{algorithm}

\medskip

Extremal bootstrap can be also applied to estimate the distribution of other statistics, including the estimators of the EV index in \eqref{eq:pickands}  or \eqref{eq:hill} and the extrapolation estimators of Section \ref{sec:extrapolation}.

\medskip

We now describe an extremal subsampling method to estimate the distributions of $Z_T(k_T)$ and $\hat{Z}_T(k_T)$ developed by \citet{chernozhukov:2011}. It is based on drawing subsamples of size $b < T$ from $(Y_1, \ldots, Y_T)$ such that $b \to \infty$ and $b/T \to 0$ as $T \to \infty$, and computing the subsampling version of the SN-statistic $Z_T(k_T)$ as 
\begin{equation}\label{eq:ssnq}
	Z^*_{b,T}(k_T) := \mathcal{A}_{b,T} (\hat{Q}_Y^{b}(\tau_b) - \hat{Q}_Y(\tau_b)), \qquad 
	\mathcal{A}_{b,T} := \frac{\sqrt{\tau_b b}}{\hat{Q}_Y^{b}(m\tau_b) - \hat{Q}^b_Y(\tau_b)},
\end{equation}
where 
$\hat{Q}_Y^{b}(\tau)$ is the sample $\tau$-quantile in the subsample of size $b$, and $\tau_b := (\tau_T T)/b$.
Similarly, the subsampling version of the  CN-Q statistic $\hat{Z}_T(k_T)$ is
\begin{equation}\label{eq:scnq}
	\hat{Z}^*_{b,T}(k_T) := \hat{A}_b (\hat{Q}_Y^{b}(\tau_b) - \hat{Q}_Y(\tau_b)),
\end{equation}
where $\hat{A}_b$ is a consistent estimator of $A_b$. For example, under the parametric restrictions specified in (\ref{univariate:param rest}), we can set $\hat{A}_b = \hat{L} b^{-\hat{\xi}}$ for $\hat{L}$ as in (\ref{univariate:L estimate}) and $\hat{\xi}$ one of the estimators in  \eqref{eq:pickands}  or \eqref{eq:hill}. 

\medskip

The distributions of $Z^*_{b,T}(k_T)$ and $\hat Z^*_{b,T}(k_T)$ over all the subsamples estimate the distributions 
of  $Z_T(k_T)$ and  $\hat Z_T(k_T)$, respectively. The number of possible subsamples depends on the structure of serial dependence in the data, but it can be very large.  In practice, the distributions over all subsamples are approximated by the distributions over a smaller number $S$ of randomly chosen subsamples, such that $S \to \infty$ as $T \to \infty$ \cite[Chap.\ 2.5]{politis:1999} . \citet{politis:1999} and \citet{bertail:2004} provide methods for choosing the subsample size $b$. 
Algorithm \ref{alg:es} describes the implementation of the extremal subsampling.

\medskip

\begin{algorithm}[Extremal Subsampling]\label{alg:es} (1) Choose the quantile index of interest $\tau_T$, the subsample size $b$ (e.g., $b =  \lfloor 50 + \sqrt{T} \rfloor$), and the number of subsamples $S$ (e.g., $S=500$).\footnote{\citet{chernozhukov:2005} suggest the choice  $b = \lfloor m + T^{1/c} \rfloor$ with $c \geq 2$ and $m>0$, to guarantee that the minimal subsample size is $m$.} (2) If the data have serial dependence,  draw $S$ subsamples from $(Y_1,  \ldots, Y_T)$ of size $b$ of the form $(Y_{1,s}^*, \ldots, Y_{b,s}^*) = (Y_k, \ldots, Y_{k+b+1})$ where $k \in \{1, \ldots, T-b+1\}$. If the data are independent, $(Y_{1,s}^*, \ldots, Y_{b,s}^*)$ can be drawn as a random subsample of size $b$  from $(Y_1, \ldots, Y_{T})$ without replacement. (3) For each $s \in \{1, \ldots, S\}$, compute $Z^*_{b,s}(k_T) $ or $\hat Z^*_{b,s}(k_T) $ applying \eqref{eq:ssnq}  or \eqref{eq:scnq} to the subsample $(Y_{1,s}^*, \ldots, Y_{b,s}^*)$. (4) Estimate the quantiles of $Z_T(k_T)$ or $\hat Z_T(k_T)$ by the sample quantiles of $\{Z^*_{b,s}(k_T)\}_{s=1}^S$ or $\{\hat Z^*_{b,s}(k_T)\}_{s=1}^S$.
\end{algorithm}

\medskip

Extremal subsampling differs from conventional subsampling, which is inconsistent for extremal quantiles. This difference can be more clearly appreciated in the case of $\hat{Z}_T(k_T)$. Here,  conventional subsampling would recenter the subsampling version of the statistic by the estimator in the full sample $\hat{Q}_Y(\tau_T)$.
Recentering in this way requires $A_b/A_T \to 0$ for consistency (see Theorem 2.2.1 in \citet{politis:1999}), but $A_b/A_T \to \infty$ when $\xi > 0$.
Thus, when $\xi > 0$ the extremal sample quantiles $\hat{Q}_Y(\tau_T)$ diverge rendering conventional subsampling to be inconsistent. In contrast, extremal subsampling uses $\hat{Q}_Y(\tau_b)$ for recentering.
This sample quantile may diverge, but because it is an intermediate order quantile if $b/T \to 0$, the speed of its divergence is strictly slower than that of $A_T$. Hence, extremal subsampling exploits the special structure of the order statistics to do the recentering. 

\medskip

\subsubsection{Median Bias Correction and Confidence Intervals}

 \citet{chernozhukov:2011} construct asymptotically median-unbiased estimators and $(1-\alpha)$-confidence intervals (CI) for $Q_Y(\tau_T)$ based on the SN-Q statistic as
\[
	\hat{Q}_Y(\tau) - \frac{\hat{c}_{1/2}}{\mathcal{A}_T} \qquad \text{and} \qquad
	\biggl[ \hat{Q}_Y(\tau) - \frac{\hat{c}_{1-\alpha/2}}{\mathcal{A}_T}, \ \hat{Q}_Y(\tau) - \frac{\hat{c}_{\alpha/2}}{\mathcal{A}_T} \biggr],
\]
where $\hat{c}_p$ is a consistent estimator of the $p$-quantile of $Z_T(k_T)$ that can be obtained using  Algorithms \ref{alg:eb} or \ref{alg:es}. \citet{chernozhukov:2011} also construct asymptotically median-unbiased estimators and $(1-\alpha)$-CIs for $Q_Y(\tau_T)$ based on the CN-Q statistic as
\[
	\hat{Q}_Y(\tau) - \frac{\hat{c}'_{1/2}}{\hat{A}_T} \qquad \text{and} \qquad
	\biggl[ \hat{Q}_Y(\tau) - \frac{\hat{c}'_{1-\alpha/2}}{\hat{A}_T}, \ \hat{Q}_Y(\tau) - \frac{\hat{c}'_{\alpha/2}}{\hat{A}_T} \biggr],
\]
where $\hat{c}'_p$ is a consistent estimator of the $p$-quantile of $\hat Z_T(k_T)$ that can be obtained using  Algorithm  \ref{alg:es}, and $\hat{A}_T$ is a consistent estimator of $A_T$ such as \eqref{univariate:A estimator}.

\medskip

\subsubsection{Extrapolation Estimator for Very Extremes}\label{sec:extrapolation}

Sample $\tau$-quantiles can be  very inaccurate estimators of marginal $\tau$-quantiles when $\tau T$ is very small, say $\tau T < 1$. For such very extremal cases we can construct more precise estimators using the assumptions on the behavior of the tails.  In particular, we can estimate less extreme quantiles reliably, and extrapolate them to the quantile of interest using the tail assumptions. 

\medskip

\citet{Dekkers:dehaan} develop the following extrapolation estimator:
\begin{equation} \label{extrapolate1}
	\tilde{Q}_{Y}(\tau_T)  = \hat{Q}_Y(\tilde \tau_T) + \frac{(\tau_T/\tilde \tau_T)^{-\hat{\xi}}-1}{2^{-\hat{\xi}}-1} \Bigl[ \hat{Q}_Y(2 \tilde \tau_T) - \hat{Q}_Y(\tilde \tau_T) \Bigr],
\end{equation}
where $\tau_T \ll \tilde \tau_T$ and $\hat \xi$ is a consistent estimator of $\xi$ such as \eqref{eq:pickands}  or \eqref{eq:hill}. Then, for $\tilde \tau_T T \to \tilde k$ and $\tau_T T \to k$ with $\tilde{k} > k$,
\[
	\frac{\tilde{Q}_Y(\tau_T) - Q_Y(\tau_T)}{\hat{Q}_Y(\tilde \tau_T) - \hat{Q}_Y(2\tilde \tau_T)} \ \to_d \ \frac{(\tilde k/k)^\xi - 2^{-\xi}}{1-2^{-\xi}} + \frac{1-(\Gamma_{\tilde{k}}/k)^\xi}{e^{\xi \mathcal{E}_{\tilde{k}}}-1},
\]
where $\mathcal{E}_{\tilde{k}}$ and $\Gamma_{\tilde{k}}$ are independent, $\Gamma_{\tilde{k}}$ has a standard gamma distribution with shape parameter $(2\tilde{k}+1)$, and $\mathcal{E}_{\tilde{k}} \sim \sum_{j=\tilde{k}+1}^{2\tilde{k}} Z_j/j$ with $Z_1, Z_2, \dots$ i.i.d.~standard exponential.  \citet{he:2016} proposed the closely related estimator
 \begin{equation} \label{extrapolate2}
	\breve{Q}_Y(\tau_T) = \hat{Q}_Y(\tilde \tau_T) + \frac{(\tau_T/\tilde \tau_T)^{-\hat{\xi}}-1}{2^{\hat{\xi}}-1} \Bigl[ \hat{Q}_Y(\tilde \tau_T/2) - \hat{Q}_Y(\tilde \tau_T) \Bigr].
\end{equation}
Under some regularity conditions, they show that for $\tau_T/\tilde \tau_T \to 0$ as $T \to \infty,$ this estimator converges to a normal distribution jointly with the EV index estimator $\hat \xi$.

\medskip

The estimators in (\ref{extrapolate1}) and (\ref{extrapolate2})  have good properties provided that the quantities on the right-hand side are well estimated, which in turn requires that $\tilde \tau_T T$ be large, and that the Pareto-type tail model be a good approximation.

\subsection{Multivariate Case: Conditional Quantiles}
The $\tau$-quantile regression ($\tau$-QR) estimator of the conditional $\tau$-quantile $Q_Y(\tau | x) = x'\beta(\tau)$ is:
\begin{equation} \label{regression}
	\hat{Q}_Y(\tau | x) := x' \hat{\beta}(\tau), \qquad \hat{\beta}(\tau) \in \argmin_{\beta \in \mathbb{R}^d} \, \sum_{t=1}^T \rho_\tau(Y_t - X_t' \beta).
\end{equation}
This estimator was introduced by \citet{laplace:1818} for the median case, and extended by \citet{koenker:1978} to include other quantiles and regressors.


\medskip

In this section, we review the asymptotic behavior of the QR estimator under extreme and intermediate order sequences, and describe inference methods for extremal quantile regression. The analysis for the multivariate case parallels the analysis for the univariate case in Section \ref{sec:univariate}.

\medskip

\subsubsection{Extreme Order Approximation} \label{regression:extreme}

Consider the {\em canonically-normalized quantile regression (CN-QR) statistic}
\begin{equation} \label{CNQR}
	\hat{Z}_T(k_T) := A_T (\hat{\beta}(\tau_T) - \beta(\tau_T)) \quad \text{for} \quad A_T := 1/Q_U(1/T),
\end{equation}
and the {\em self-normalized quantile regression (SN-QR) statistic}
\begin{equation} \label{SNQR}
	Z_{T}(k_T) := \mathcal{A}_T (\hat{\beta}(\tau_T) - \beta(\tau_T)) \quad \text{for} \quad \mathcal{A}_T := \frac{\sqrt{k_T}}{\bar{X}_T'(\hat{\beta}(m\tau_T)-\hat{\beta}(\tau_T))},
\end{equation}
where $\bar{X}_T := T^{-1} \sum_{t=1}^T X_t$ and $m$ is a real number such that $k (m-1) > d_x$ for $k_T = \tau_T T \to k$. For example, $m = (d_x+p)/k_T + 1 = (d_x+p)/k + 1 + o(1)$ where $p \geq 1$ is a spacing parameter (e.g.\ $p = 5$).
The CN-QR statistic is generally infeasible for inference because it depends on the unknown canonical normalization constant $A_T$. This constant can only be estimated consistently under strong parametric assumptions, which will be discussed in Section \ref{regression:A}. The SN-QR statistic is always feasible because  it  uses a  normalization that only depends on the data.

\medskip

\citet{chernozhukov:2011} show that for $k_T \to k > 0$ as $T \to \infty$,
\begin{equation} \label{reg EV}
	\hat{Z}_T(k_T) \ \to_d \ \hat{Z}_\infty(k),
\end{equation}
where for $\chi = 1$ if $\xi < 0$ and $\chi = -1$ if $\xi > 0$,
\begin{equation} \label{CNQRlimit}
	\hat{Z}_\infty(k) := \chi \cdot \argmin_{z \in \mathbb{R}^{d_x}} \, \Biggl[ -kE[X]'z + \sum_{t=1}^{\infty} \{\mathcal{X}_t'z - \chi (\Gamma_t^{-\xi} - k^{-\xi}) \mathcal{X}_t' \gamma \}_+ \Biggr],
\end{equation}
where $\{\mathcal{X}_1, \mathcal{X}_2, \dots\}$ is an i.i.d.\ sequence with distribution $F_X$; $\{\Gamma_1, \Gamma_2, \dots\} := \{\mathcal{E}_1, \mathcal{E}_1+\mathcal{E}_2, \dots\}$; $\{\mathcal{E}_1, \mathcal{E}_2, \dots\}$ is an i.i.d.\ sequence of standard exponential variables that is independent of $\{\mathcal{X}_1, \mathcal{X}_2, \dots\}$; and $\{y\}_+ := \max(0,y)$.
Furthermore,
\begin{equation} \label{reg EV2}
	Z_T(k_T) \ \to_d \ Z_\infty(k) := \frac{\sqrt{k} \hat{Z}_\infty(k)}{E[X]' (\hat{Z}_\infty(mk) - \hat{Z}_\infty(k)) + \chi \cdot (m^{-\xi} - 1) k^{-\xi}}.
\end{equation}
The limit EV distributions are more complicated than in the univariate case, but they share some common features.
First, they depend crucially on the gamma variables $\Gamma_t$, are not necessarily centered at zero, and can have a significant first-order asymptotic median bias. Second, as mentioned above, the limit distribution of the CN-QR statistic in Equation (\ref{CNQRlimit}) is generally infeasible for inference due to the difficulty in consistently estimating the scaling constant $A_T$.

\begin{remark}[Very Extreme Order Quantiles]
 \citet{feigin:1994},  \citet{smith:1994}, \citet{chernozhukov:1998}, \citet{portnoy:jur}, and \citet{knight:linear}  derived related results for canonically normalized linear programing or frontier regression estimators under very extreme order sequences where $\tau_T T \searrow 0$ as $T \to \infty$.
\end{remark}

\medskip

\subsubsection{Intermediate Order Approximation} \label{regression:intermediate}

\citet{victor:annals} shows that for $\tau_T \searrow 0$ and $k_T \to \infty$ as $T \to \infty$,
\begin{equation} \label{reg normal}
	Z_T(k_T) =  \mathcal{A}_T (\hat{\beta}(\tau_T) - \beta(\tau_T)) \ \to_d \ \mathcal{N} \biggl( 0, \, E[XX']^{-1} \frac{\xi^2}{( m^{-\xi}-1)^2 } \biggr),
\end{equation}
where $\mathcal{A}_T$ is defined as in (\ref{SNQR}). As in the univariate case, this normal approximation provides a less accurate approximation to the distribution of the extremal quantile regression than the EV approximation  when $k_T \not\to \infty$. The condition  $k_T \to \infty$ can be interpreted in finite samples as requiring that $k_T/d_x \geq 30$, where $k_T/d_x$ is a dimension-adjusted order of the quantile explained in Section \ref{thumb}.

\medskip

\subsubsection{Estimation of $\xi$ and $\gamma$} \label{regression:ev index}

Some inference methods for extremal quantile regression require consistent estimators of the EV index $\xi$ and the scale parameter $\gamma$.
The regression analog of the Pickands estimator is
%
\begin{equation} \label{regression:pickands}
	\hat{\xi}_P := \frac{-1}{\ln 2} \ln \biggl( \frac{\bar{X}_T' (\hat{\beta}(4\tilde{\tau}_T) - \hat{\beta}(2\tilde{\tau}_T))}{\bar{X}_T' (\hat{\beta}(2\tilde{\tau}_T) - \hat{\beta}(\tilde{\tau}_T))} \biggr).
\end{equation}
This estimator is consistent if $\tilde{\tau}_T T \to \infty$ and $\tilde{\tau}_T \searrow 0$ as $T \to \infty$.
Under additional regularity conditions, for $\tilde{\tau}_T \searrow 0$ and $\tilde{\tau}_T T \to \infty$ as $T \to \infty$,
\begin{equation} \label{regression:pickands:dist}
	\sqrt{\tilde{\tau}_T T} (\hat{\xi}_P - \xi) \ \to_d \ \mathcal{N} \biggl( 0, \, \frac{\xi^2(2^{2\xi+1}+1)}{[2 (2^{\xi}-1) \ln 2]^2} \biggr).
\end{equation}
The regression analog of the Hill estimator is
\begin{equation} \label{regression:hill}
	\hat{\xi}_H := -\frac{\sum_{t=1}^T 1\{Y_t<X_t'\hat{\beta}(\tilde{\tau}_T)\} \ln(Y_t/X_t'\hat{\beta}(\tilde{\tau}_T))}{\tilde{\tau}_T\sum_{t=1}^T 1\{Y_t<X_t'\hat{\beta}(\tilde{\tau}_T)\}},
\end{equation}
which is applicable when $\xi>0$ and $X_t\hat{\beta}(\tilde{\tau}_T)<0$.
Under further regularity conditions, for $\tilde{\tau}_T \searrow 0$ and $\tilde{\tau}_T T \to \infty$,
\begin{equation} \label{regression:hill:dist}
	\sqrt{\tilde{\tau}_T T} (\hat{\xi}_H-\xi) \ \to_d \ \mathcal{N}(0,\xi^2).
\end{equation}
These limit results can be used to construct confidence intervals for $\xi$.
The scale parameter $\gamma$ can be estimated by
\begin{equation} \label{regression:scale}
	\hat{\gamma} = \frac{\hat{\beta}(2\tilde{\tau}_T) - \hat{\beta}(\tilde{\tau}_T)}{\bar{X}_T' (\hat{\beta}(2\tilde{\tau}_T) - \hat{\beta}(\tilde{\tau}_T))},
\end{equation}
which is consistent if $\tilde{\tau}_T T \to \infty$ and $\tilde{\tau}_T \searrow 0$ as $T \to \infty$.

\medskip

The choice of  $\tilde{\tau}_T$ is similar to the univariate case in  Section \ref{univariate:ev index}. This time, however, one needs to take into account the multivariate nature of the problem. For example, if the interest lies in making the inference on $\beta(\tau_T)$ for a particular $\tau_T$, it is reasonable to set $\tilde{\tau}_T$ equal to the closest value to $\tau_T$ such that  $\tilde{\tau}_T T/d_x \geq 30$. We refer again the reader to Section \ref{thumb} for a  discussion on the difference in the choice of $\tilde{\tau}_T$ between the univariate and multivariate cases.

\medskip

\subsubsection{Estimation of $A_T$} \label{regression:A}
To use the CN-QR statistic for inference, we need to estimate the scaling constant $A_T$ defined in \eqref{CNQR}. This requires strong restrictions and an additional estimation procedure.
For example, assume that the non-parametric slowly varying component $L(\tau)$ of $A_T$ is replaced by a constant $L$, i.e., as $\tau \searrow 0$
\begin{equation} \label{regression:param rest}
	1/Q_U(\tau) = L \cdot \tau^{\xi} \cdot (1 + \delta(\tau)) \quad \text{for some $L \in \mathbb{R}$ and $\delta(\tau) \to 0$.}
\end{equation}
Then we can estimate the constant $L$ by
\begin{equation} \label{regression:L estimate}
	\hat{L} := \frac{\bar{X}_T' (\hat{\beta}(2\tilde{\tau}_T) - \hat{\beta}(\tilde{\tau}_T))}{(2^{-\hat{\xi}}-1) \cdot \tilde{\tau}_T^{-\hat{\xi}}},
\end{equation}
where $\hat{\xi}$ is either the Pickands or Hill estimator given in \eqref{regression:pickands} or \eqref{regression:hill}.
Thus, the scaling constant $A_T$ is estimated by $$\hat{A}_T := \hat{L} T^{-\hat{\xi}}.$$

\subsubsection{Computing Quantiles of the Limit EV Distributions}

We consider inference and asymptotically median unbiased estimation for linear functions  of the coefficient vector $\beta(\tau)$, 
$\psi' \beta(\tau),$
for some nonzero vector $\psi \in \mathbb{R}^{d_x}$, based on the EV approximations $\psi'\hat Z_{\infty}(k)$ from  \eqref{reg EV2} and $\psi' Z_{\infty}(k)$ from \eqref{reg EV2}. We describe three methods to  compute critical values of the limit EV distributions:  analytical computation, extremal bootstrap, and extremal subsampling. The analytical and bootstrap methods require estimation of the EV index $\xi$ and the scale parameter $\gamma$. 
Subsampling applies under more general conditions than the other methods, and hence we would recommend the use of it.
However, the analytical and  bootstrap methods can be more accurate than subsampling if the data satisfy a stronger set of assumptions.

\medskip

The analytical computation method is based directly on the limit distributions \eqref{reg EV} and \eqref{reg EV2} replacing $\xi$ and $\gamma$ by  consistent estimators. Define the $d_x$-dimensional random vector:
\begin{equation} \label{analytical}
	\hat{Z}^\ast_\infty(k) = \hat{\chi} \cdot \argmin_{z \in \mathbb{R}^{d_x}} \, \Biggl[ -k \bar{X}_T' z + \sum_{t=1}^\infty \{\mathcal{X}_t' z - \hat{\chi} (\Gamma_t^{-\hat{\xi}} - k^{-\hat{\xi}}) \mathcal{X}_t' \hat{\gamma}\}_+ \Biggr],
\end{equation}
where $\hat{\chi} = 1$ if $\hat{\xi} < 0$ and $\hat{\chi} = -1$ if $\hat{\xi} > 0$,  $\hat{\xi}$ is an estimator of $\xi$ such as \eqref{regression:pickands} or \eqref{regression:hill},  $\hat{\gamma}$  is an estimator of $\gamma$ such as  \eqref{regression:scale}, $\{\Gamma_1, \Gamma_2, \dots\} = \{\mathcal{E}_1, \mathcal{E}_1 + \mathcal{E}_2, \dots\}$, $\{\mathcal{E}_1, \mathcal{E}_2, \dots\}$ is an i.i.d.\ sequence of standard exponential variables, and $\{\mathcal{X}_1, \mathcal{X}_2, \dots\}$ is an i.i.d.\ sequence independent of $\{\mathcal{E}_1, \mathcal{E}_2, \dots\}$ with distribution function $\hat{F}_X$, where $\hat{F}_X$ is any smooth consistent estimator of $F_X$, e.g., a smoothed empirical distribution function of the sample $(X_1, \ldots, X_T)$.
Also, let
\[
	Z^\ast_\infty(k) = \frac{\sqrt{k} \hat{Z}^\ast_\infty(k)}{\bar{X}_T'(\hat{Z}^\ast_\infty(mk)-\hat{Z}^\ast_\infty(k))+\hat{\chi}(m^{-\hat{\xi}}-1)k^{-\hat{\xi}}}.
\]
The quantiles of $\psi'\hat Z_{\infty}(k)$ and $\psi' Z_{\infty}(k)$  are estimated by  the  corresponding quantiles of the  $\psi' \hat{Z}^\ast_\infty(k)$ and $\psi' Z^\ast_\infty(k)$, respectively.
In practice, these quantiles can only be evaluated numerically via the following algorithm.

\begin{algorithm}[QR Analytical Computation]\label{alg:cac}  (1) Choose the quantile index of interest $\tau_T$ and the number of simulations $S$ (e.g., $S=200$). (2) For each $s \in \{1, \ldots, S\}$, draw an i.i.d. sequence $\{\hat Z^*_{\infty,s}(k), \dots, \hat Z^*_{\infty,s}(k)\}$ from the random vector $Z_{\infty}^*(k)$ defined in  \eqref{analytical}
 with $k = \tau_T T$ and the infinite summation truncated at some finite value $M$ (e.g. $M=T$). (3) For each $s \in \{1, \ldots, S\}$, compute $Z^\ast_{\infty,s}(k) = \sqrt{k} \hat{Z}^\ast_{\infty,s}(k)/(\bar{X}_T'(\hat{Z}^\ast_{\infty,s}(mk)-\hat{Z}^\ast_{\infty,s}(k))+\hat{\chi}(m^{-\hat{\xi}}-1)k^{-\hat{\xi}})$. (4) Estimate the quantiles of  $\psi'\hat Z_{T}(k_T)$ and $\psi' Z_{T}(k_T)$ by the sample quantiles of $\{\psi'\hat Z^*_{T,s}(k_T)\}_{s=1}^S$ and $\{\psi'Z^*_{T,s}(k_T)\}_{s=1}^S$.
%
\end{algorithm}

The extremal bootstrap is computationally less demanding than the analytical methods. It is based on simulating samples from a random variable with the same tail behavior as $(Y_1, \ldots, Y_T)$. Consider the bootstrap sample $\{(Y_1^*, X_1), \dots, (Y_T,^* X_T)\}$, where
\begin{equation} \label{generalized pareto regression}
	 Y_t^*= \frac{\mathcal{E}_t^{-\xi}-1}{-\xi} X_t' {\gamma}, \ \  \mathcal{E}_t \sim i.i.d. \ \ \text{Exponential}(1),
\end{equation}
 and $\{X_1, \dots, X_T\}$ is a fixed set of observed regressors from the data. The variable $Y_t^*$ has the conditional quantile function 
\begin{equation} \label{cquant:gp}
	Q_{Y_t^*}(\tau|x) =  x'\beta^*(\tau), \qquad \beta^*(\tau) = \frac{[-\ln (1-\tau)]^{-\xi} -1}{-\xi} \gamma,
\end{equation}
The extremal bootstrap approximates  the distribution of $Z_T(k_T) = \mathcal{A}_T (\hat{\beta}(\tau)-\beta(\tau))$ by the  distribution of $Z^*_T(k_T) = \mathcal{A}_T (\hat{\beta}^*(\tau)- \beta^*(\tau))$ where $\hat{\beta}^*(\tau)$ is the $\tau$-QR estimator in the bootstrap sample.  This approximation reproduces both the EV limit \eqref{reg EV2} under extreme value sequences, and the normal limit \eqref{reg normal} under intermediate order sequences. The distribution of $Z^*_T(k_T)$ can be obtained by simulation using the algorithm:

\begin{algorithm}[QR Extremal Bootstrap]\label{alg:ceb} (1) Choose the quantile index of interest $\tau_T$ and the number of simulations $S$ (e.g., $S=500$). (2) For each $s \in \{1, \ldots, S\}$, draw a bootstrap sample $\{(Y^*_{1,s},X_1),  \dots, (Y^*_{T,s},X_T)\}$ from the random vector $\{(Y_1^*, X_1), \dots, (Y_T,^* X_T)\}$ defined in  \eqref{generalized pareto regression}, replacing $\xi$ by the estimator  \eqref{regression:pickands} or \eqref{regression:hill}  and $\gamma$ by the estimator  \eqref{regression:scale}. (3) For each $s \in \{1, \ldots, S\}$, compute the statistic $Z^*_{T,s}(k_T) = \mathcal{A}_T (\hat{\beta}_s^*(\tau_T)- \beta^*(\tau_T))$, where $\hat{\beta}_s^*(\tau_T)$ is the $\tau_T$-QR in the bootstrap sample $\{(Y^*_{1,s},X_1),  \dots, (Y^*_{T,s},X_T)\}$ and $\beta^*(\tau_T)$ is defined as in \eqref{cquant:gp} replacing $\xi$ and $\gamma$ by the same estimators as in step (2). (4) Estimate the quantiles of $\psi'Z_T(k_T)$ by the sample quantiles of $\{\psi'Z^*_{T,s}(k_T)\}_{s=1}^S$.
\end{algorithm}

\medskip

\citet{chernozhukov:2011} developed an extremal subsampling method to estimate the distributions of $\hat Z_T(k_T)$ and $Z_T(k_T)$. It is based on drawing subsamples of size $b < T$ from $\{(X_t,Y_t)\}_{t=1}^T$ such that $b \to \infty$ and $b/T \to 0$ as $T \to \infty$, and computing the subsampling version of the SN-QR statistic as
\begin{equation}\label{eq:ssnqr}
	Z^*_{b,T}(k_T) := \mathcal{A}_{b,T} (\hat{\beta}_{b}(\tau_b) - \hat{\beta}(\tau_b)), \qquad 
	\mathcal{A}_{b,T} := \frac{\sqrt{\tau_b b}}{\bar{X}_{b,T}'[\hat{\beta}_b(m\tau_b) - \hat{\beta}_b(\tau_b)]},
\end{equation}
where $m=(d_x + p)/(\tau_T T)$ for some {\em spacing parameter} $p \geq 1$ (e.g.\ $p=5$), $\hat{\beta}_b(\tau)$ is the  $\tau$-QR estimator in the subsample of size $b$, $\bar{X}_{b,T}$ is the sample mean of the regressors in the subsample, and $\tau_b := (\tau_T T)/b$.\footnote{In practice, it is reasonable to use the following finite-sample adjustment to $\tau_b$: $\tau_b = \min\{(\tau_T T)/b, 0.2\}$ if $\tau_T < 0.2$, and $\tau_b = \tau_T$ if $\tau_T \geq 0.2$. The idea is that $\tau_T$ is adjusted to be non-extremal if $\tau_T > 0.2$, and the subsampling procedure reverts to central order inference. The truncation of $\tau_b$ by $0.2$ is a finite-sample adjustment that restricts the key statistics $Z^*_{b,T}(k_T)$ to be extremal in subsamples. These finite-sample adjustments do not affect the asymptotic arguments.} 
Similarly, the subsampling version of the  CN-QR statistic $\hat{Z}_T(k_T)$ is
\begin{equation}\label{eq:scnqr}
	\hat{Z}^*_{b,T}(k_T) := \hat{A}_b (\hat{\beta}_b(\tau_b) - \hat{\beta}(\tau_b)),
\end{equation}
where $\hat{A}_b$ is a consistent estimator for $A_b$. For example,  $\hat{A}_b = \hat{L} b^{-\hat{\xi}},$ for $\hat{L}$ given by (\ref{regression:L estimate}) and  $\hat{\xi}$ is  the estimator of $\xi$ given in \eqref{regression:pickands} or \eqref{regression:hill}.

\medskip

As in the univariate case, the distributions of $Z^*_{b,T}(k_T)$ and $\hat Z^*_{b,T}(k_T)$ over all the possible subsamples estimate the distributions  of  $Z_T(k_T)$ and  $\hat Z_T(k_T)$, respectively. These distributions can be obtained by simulation using the algorithm:
\medskip

\begin{algorithm}[QR Extremal Subsampling]\label{alg:ces} (1) Choose the quantile index of interest $\tau_T$, the subsample size $b$ (e.g., $b =  \lfloor 50 + \sqrt{T} \rfloor$), and the number of subsamples $S$ (e.g., $S=500$). (2) If the data have serial dependence,  draw $S$ subsamples from $\{(Y_t, X_t)\}_{t=1}^T$ of size $b$,  $\{(Y_{t,s}^*, X_{t,s}^*)\}_{t=1}^b$, of the form $ (Y_{t,s}^*, X_{t,s}^*) = (Y_{t+k}, X_{t+k})$ where $k \in \{1, \ldots, T-b+1\}$. If the data are independent, $\{(Y_{t,s}^*, X_{t,s}^*)\}_{t=1}^b$ can be drawn as a random subsample of size $b$  from $\{(Y_t, X_t)\}_{t=1}^T$ without replacement. (3) For each $s \in \{1, \ldots, S\}$, compute $Z^*_{b,s}(k_T) $ or $\hat Z^*_{b,s}(k_T) $ applying \eqref{eq:ssnqr}  or \eqref{eq:scnqr} to the subsample $\{(Y_{t,s}^*, X_{t,s}^*)\}_{t=1}^b$. (4) Estimate the quantiles of $\psi'Z_T(k_T)$ or $\psi'\hat Z_T(k_T)$ by the sample quantiles of $\{\psi'Z^*_{b,s}(k_T)\}_{s=1}^S$ or $\{\psi'\hat Z^*_{b,s}(k_T)\}_{s=1}^S$.
\end{algorithm}

\medskip

The comments of  Section \ref{univariate:cv}  on the choice of subsample size, number of simulations, and differences with conventional subsampling also apply to the regression case.

\medskip


\subsubsection{Median Bias Correction and Confidence Intervals}
\citet{chernozhukov:2011}  construct asymptotically median-unbiased estimators and $(1-\alpha)$-CIs for $\psi'\beta(\tau)$  based on the SN-QR statistic as 
\[
	\psi' \hat{\beta}(\tau) - \frac{\hat{c}_{1/2}}{\mathcal{A}_T} \qquad \text{and} \qquad
	\biggl[ \psi' \hat{\beta}(\tau) - \frac{\hat{c}_{1-\alpha/2}}{\mathcal{A}_T}, \ \psi' \hat{\beta}(\tau) - \frac{\hat{c}_{\alpha/2}}{\mathcal{A}_T} \biggr],
\]
where  $\hat{c}_p$ is a  consistent estimator of the $p$-quantile $c_\alpha$ of $Z_T(k_T)$ that can be obtained using Algorithms \ref{alg:cac}, \ref{alg:ceb}, or \ref{alg:ces}. \citet{chernozhukov:2011}  also construct asymptotically median-unbiased estimators and $(1-\alpha)$-CIs for $\psi'\beta(\tau)$  based on the CN-QR statistic as
\[
	\psi' \hat{\beta}(\tau) - \frac{\hat{c}_{1/2}'}{A_T} \qquad \text{and} \qquad
	\biggl[ \psi' \hat{\beta}(\tau) - \frac{\hat{c}_{1-\alpha/2}'}{A_T}, \ \psi' \hat{\beta}(\tau) - \frac{\hat{c}_{\alpha/2}'}{A_T} \biggr],
\]
where $\hat{c}_p'$ is a consistent estimator of the $p$-quantile of $\hat{Z}_T(k_T)$ that can be obtained using Algorithms \ref{alg:cac} or \ref{alg:ces}.

\medskip

\subsubsection{Extrapolation Estimator for Very Extremes} The $\tau$-QR estimators can be very inaccurate  when $\tau T/d_x$ is very small, say $\tau T/d_x < 1$. We can construct extrapolation estimators  for these cases that use the assumptions on the behavior of the tails. By analogy with the univariate case, 
\begin{equation} \label{regression:extrapo1}
	\tilde{\beta}(\tau_T) = \hat{\beta}(\tilde \tau_T) + \frac{(\tau_T / \tilde \tau_T)^{-\hat{\xi}}-1}{2^{-\hat{\xi}}-1} \bigl[ \hat{\beta}(2\tilde \tau_T) - \hat{\beta}(\tilde \tau_T) \bigr],
\end{equation}
or
\begin{equation} \label{regression:extrapo2}
	\breve{\beta}(\tau_T) = \hat{\beta}(\tilde \tau_T) + \frac{(\tau_T/\tilde \tau_T)^{-\hat{\xi}}-1}{2^{\hat{\xi}}-1} \Bigl[ \hat{\beta}(\tilde \tau_T/2) - \hat{\beta}(\tilde \tau_T) \Bigr],
\end{equation}
where  $\tau_T \ll \tilde \tau_T$, and $\hat \xi$ is the Pickands or Hill estimator of $\xi$ in \eqref{regression:pickands} or \eqref{regression:hill}. \citet{he:2016} derived the joint asymptotic distribution of   $(\breve{\beta}(\tau_T),\hat{\xi}_P)$. \citet{wang:2012}  developed other  extrapolation estimators  for heavy-tailed distributions with $\xi > 0$.

\medskip

The estimators in (\ref{regression:extrapo1}) and (\ref{regression:extrapo2})  have good properties provided that the quantities on the right-hand side are well estimated, which in turn requires that $\tilde \tau_T T/d_x$ be large, and that the Pareto-type tail model be a good approximation.
To construct the confidence interval for $\beta(\tau_T)$ based on extrapolation, we can apply the extremal subsampling to the statistic 
$$\widetilde{\mathcal{A}}_T [\tilde{\beta}(\tau_T) - \beta(\tau_T)], \quad \widetilde{\mathcal{A}}_T =  \frac{\sqrt{\tilde \tau_T T}}{\bar{X}_T'(\hat{\beta}(m\tilde \tau_T)-\hat{\beta}(\tilde \tau_T))}.$$ 
For the estimator (\ref{regression:extrapo2}), we can also use analytical methods based on the asymptotic distribution given in Corollary 3.4 of  \citet{he:2016}.

\medskip

\subsection{EV Versus Normal Inference} \label{thumb}

 \citet{chernozhukov:2011} provided a simple rule of thumb for the application of EV inference.
Recall that the order of a sample $\tau_T$-quantile from a sample of size $T$ is  $\tau_T T$ (rounded to the next integer). This order plays a crucial role in determining the quality of the EV or normal approximations.
Indeed, the former requires $\tau_T T \to k$, whereas the latter requires $\tau_T T \to \infty$.
In the regression case, in addition to the order of the quantile, we need to take into account $d_x$, the dimension of $X$. As an example, consider the case where all $d_x$ covariates are indicators that  divide equally the sample into subsamples of size $T/d_x$. Then, each of the components of the $\tau_T$-QR estimator will correspond to a sample quantile of order $\tau_T T/d_x$.
We may therefore think of  $\tau_T T/d_x$ as a dimension-adjusted order for quantile regression.

\medskip

A common simple rule for the application of the normal is that the sample size is greater than 30.
This suggests that we should use extremal inference whenever $\tau_T T/d_x \lesssim 30$.
This simple rule may or may not be conservative.
For example, when regressors are continuous, the computational experiments in  \citet{chernozhukov:2011}  show that the normal inference performs as well as the EV inference provided that $\tau_T T/d_x \gtrsim 15$ to $20$, which suggests using EV inference when $\tau_T T/d_x \lesssim 15$ to $20$ for this case.
On the other hand, if we have an indicator in $X$ equal to one only for 2\% of the sample, then the coefficient of this indicator behaves as a sample quantile of order  $.02 \tau_T T = \tau_T T/ 50$, which would motivate using EV inference when $\tau_T T/50 \lesssim 15$ to $20$ in this case.
This rule is far more conservative than the original simple rule when $d_x \ll 50$.
Overall, it seems prudent to use both EV and normal inference methods in most cases, with the idea that the discrepancies between the two can indicate extreme situations.


\section{Empirical Applications} \label{sec:empirics}
We consider two applications of extremal quantile regression to conditional value-at-risk and financial contagion. We implement the empirical analysis in \verb"R" language with Koenker (\citeyear{koenker:2016}) \verb"quantreg" package  and the code from \citet{chernozhukov-du:2008} and \citet{chernozhukov:2011}.
The data are obtained from Yahoo! Finance.\footnote{The dataset and the code are available online at Fern\'andez-Val's website: \texttt{http://sites.bu.edu/ivanf/research/}.}

\subsection{Value-at-Risk Prediction}

We revisit the problem of forecasting the conditional value-at-risk of a financial institution posed by \citet{vl} with more recent methodology. The response variable $Y_t$ is the daily return of the Citigroup stock, and the covariates $X_{1t}$, $X_{2t}$, and $X_{3t}$ are the lagged daily returns of the Citigroup stock (C), the Dow Jones Industrial Index (DJI), and the Dow Jones US Financial Index (DJUSFN), respectively. The lagged own return captures dynamics, DJI is a measure of overall market return, and the DJUSFN is a measure of market return in the financial sector. 
We estimate quantiles of $Y_t$ conditional on $X_t = (1, X_{1t}^+, X_{1t}^-, X_{2t}^+, X_{2t}^-, X_{3t}^+, X_{3t}^-)$ with $x^+ = \max\{x, 0\}$ and $x^- = -\min\{x, 0\}$.
There are 1,738 daily observations in the sample covering the period from January 1, 2009 to November 30, 2015.

\medskip

\begin{figure}
\includegraphics[width=\textwidth]{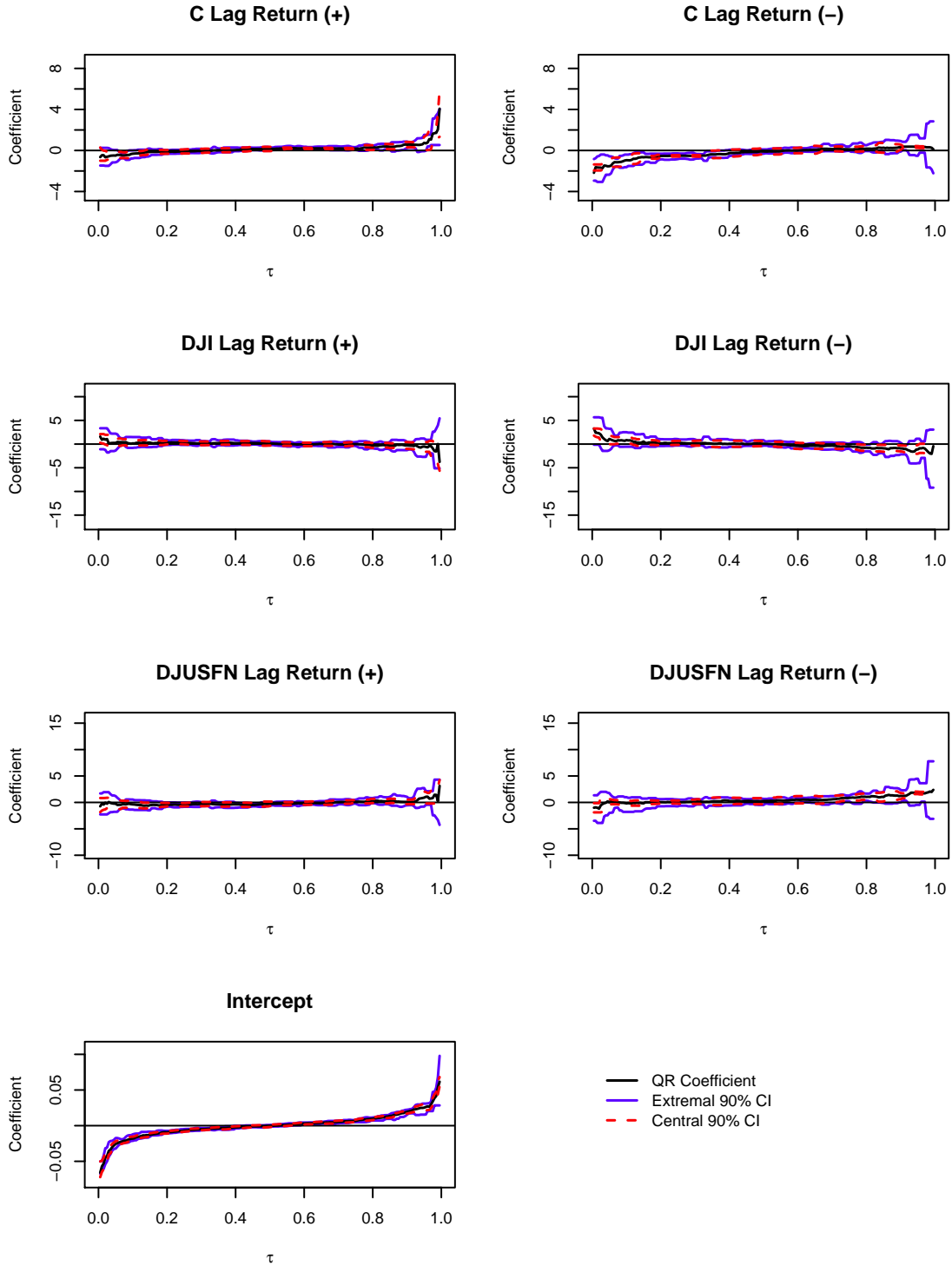}
\caption{Value-at-Risk: QR coefficient estimates and 90\% pointwise CIs. The response variable is the daily Citigroup return  from January 1, 2009 to November 30, 2015. The solid lines depict extremal CIs and the dashed lines depict normal CIs.} \label{fig1-1}
\end{figure}

\begin{figure}
\includegraphics[width=\textwidth]{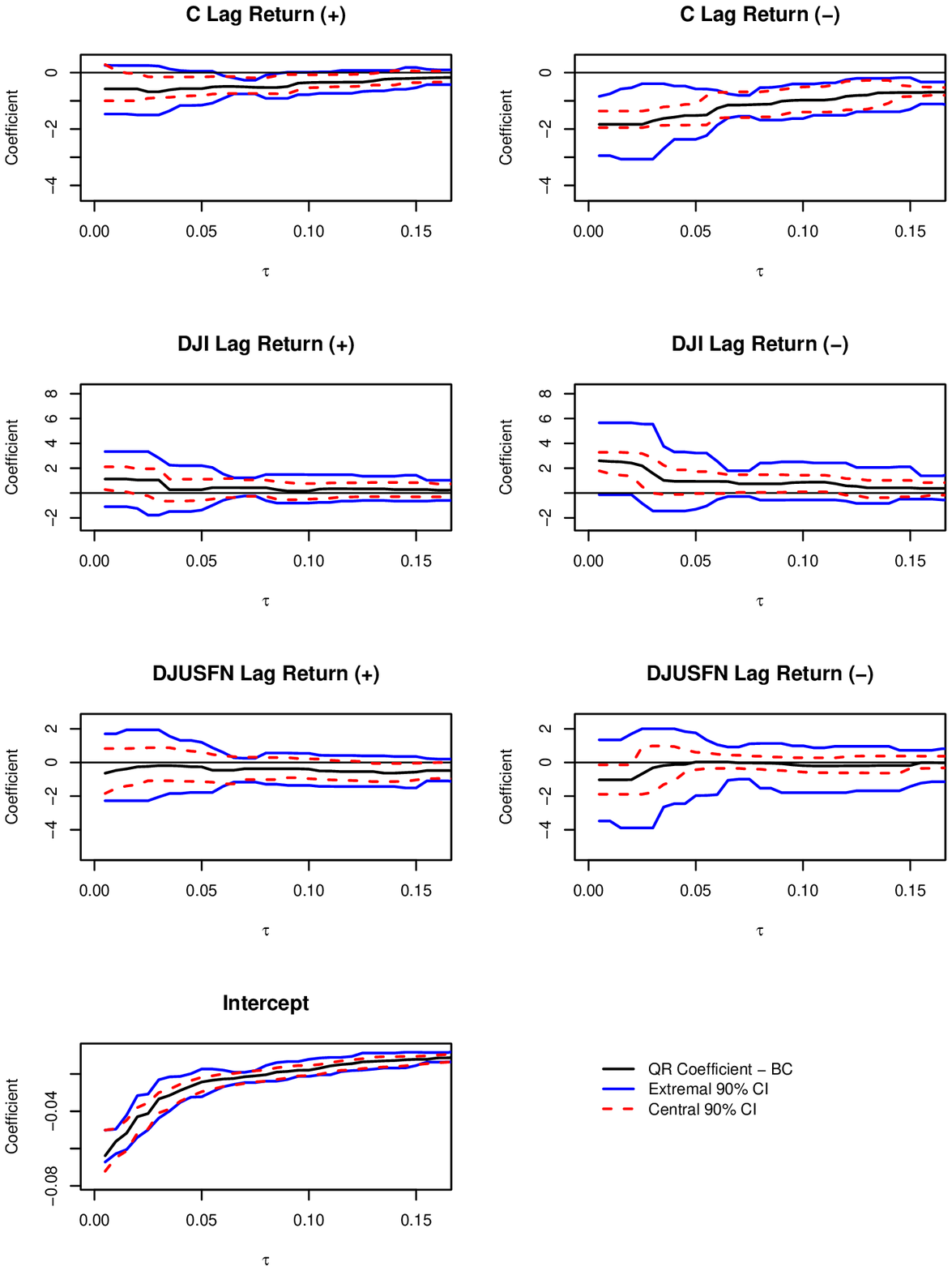}
\caption{Value-at-Risk: Bias-corrected QR coefficient estimates and 90\% pointwise CIs for low quantiles. The response variable is the daily Citigroup return  from January 1, 2009 to November 30, 2015. The solid lines depict extremal CIs and the dashed lines depict normal CIs.} \label{fig1-2}
\end{figure}

Figure \ref{fig1-1} plots the QR estimates $\hat{\beta}(\tau)$ along with 90\% pointwise CIs.
The solid lines represent the extremal CIs and the dashed lines the normal CIs.
The extremal CIs are computed by the extremal subsampling method described in Algorithm \ref{alg:ces} with the subsample size $b = \lfloor 50 + \sqrt{1,738} \rfloor = 91$  and the number of simulations $S = 500$. We use the SN-QR statistic with spacing parameter $p=5$. 
The normal CIs are based on the normal approximation with the standard errors computed with the method proposed by \citet{powell:1991}.\footnote{We used  the command \texttt{summary.rq} with the option \texttt{ker}  in the \texttt{quantreg} package to compute the standard errors.}
Figure \ref{fig1-2} plots the median bias-corrected QR estimates along with 90\% pointwise CIs for the lower tail (note that due to the median bias-correction, the coefficient estimates are slightly different from Figure \ref{fig1-1}). The bias correction is also implemented using extremal subsampling with the same specifications.


\medskip

We focus the discussion on the impact of downward movements of the explanatory variables (the C lag $X_{1t}^-$, the DJI lag $X_{2t}^-$, and the DJUSFN lag $X_{3t}^-$) on the extreme risk, that is, on the low conditional quantiles of the Citigroup stock return.
To interpret the results, it is helpful to keep in mind that if the covariates were completely irrelevant (i.e. independent from the response), then their coefficients would be equal to zero uniformly over $\tau$, except for the constant term. The intercept would coincide with the unconditional quantile of Citigroup daily return.
Another general remark is that we would expect the estimates and CIs  to be more volatile at the tails than at the center due to data sparsity. Figures \ref{fig1-1} and \ref{fig1-2} show that most of the coefficients are insignificant throughout the distribution, what confirms the expected unpredictability of the stock returns.
However, we do find that the coefficient on the Citigroup's lagged return $X_{1t}^-$ is significantly different from zero in the extreme low quantiles (see the upper right figure in Figure \ref{fig1-2}).
This suggests that from 2009 to 2015, a past drop in the stock price of Citigroup has significantly pushed down the extreme low quantiles of the current stock price.
Informally speaking, the negative return on the stock price induced the risk of a further negative outcome in the near future.

\medskip

Comparing the CIs produced by the extremal inference and the normal inference, 
Figure \ref{fig1-1} shows that they closely match in the central region, while Figures \ref{fig1-2} reveals that the normal CIs are often narrower than the extremal CIs in the tails, especially for $\tau < 0.05$.
As briefly mentioned in Section \ref{univariate:intermediate}, the extremal CIs coincide with the normal CIs when the situation is non-extremal.
Therefore, this discrepancy indicates that the normal CIs on the tails substantially underestimates the sampling variation and hence it might lead to a substantial undercoverages in the CIs.

\medskip

We next characterize the tail properties of the model.
Table \ref{table1-1} reports the estimates of the EV index $\xi$ obtained by the Hill estimator in (\ref{regression:hill}), together with bias corrected estimates and  90\% CIs based on (\ref{regression:hill:dist}), which were obtained using the QR extremal bootstrap of Algorithm \ref{alg:ceb} with $S=500$ applied to the Hill estimator.
The bias-corrected estimates of $\xi$ are relatively stable even at the extreme tails. They are greater than zero, confirming that the distribution of stock returns has a much thicker lower tail than the normal distribution. It is noteworthy that none of these  estimates were used to produce the fig.  \ref{fig1-1} and  \ref{fig1-2} because they were obtained from extremal subsampling method applied to the SN-QR statistic.

\begin{table}[hptb]
\caption{Value at Risk: Hill Estimation Results for the EV Index} \label{table1-1}
\begin{tabular}{lccc}
\hline \hline
& Estimate & Bias-Corrected & 90\% Confidence \\
& & Estimate & Interval \\
\hline
$\tau = 0.01$ & $0.330$ & $0.311$ & $[0.164, \ 0.441]$ \\
$\tau = 0.05$ & $0.427$ & $0.350$ & $[0.250, \ 0.461]$ \\
$\tau = 0.1$ & $0.447$ & $0.293$ & $[0.215, \ 0.369]$ \\
\hline\hline
\end{tabular}
\end{table}

\medskip

\begin{figure}
\includegraphics[width=\textwidth]{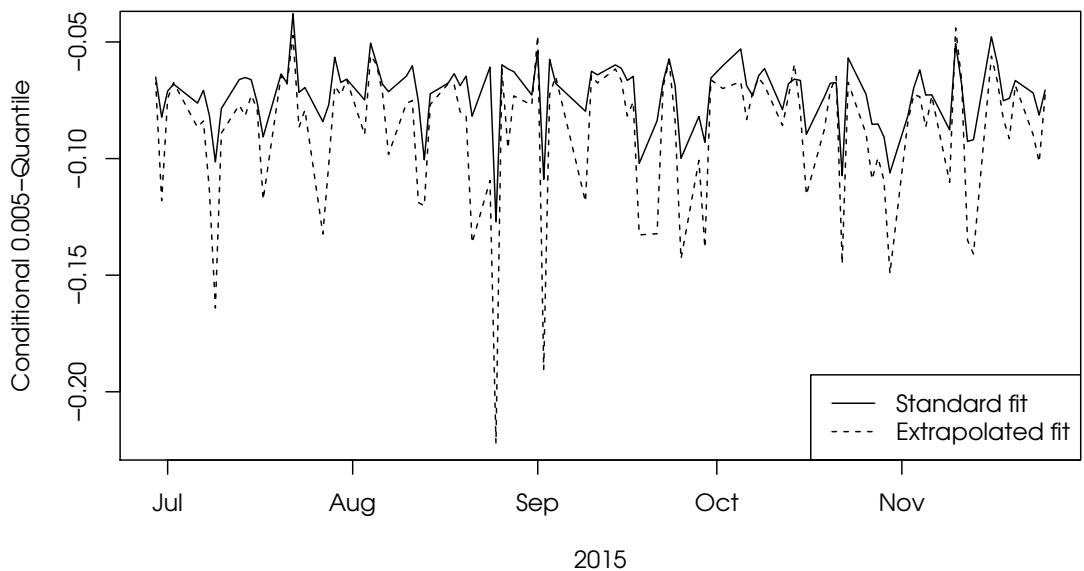}
\caption{Value-at-Risk: Extrapolation vs Standard Estimates of Conditional Quantiles of the daily Citigroup return between July 1, 2015 and November 30, 2015.} \label{fig1-4}
\end{figure}

Having characterized the EV index, we can now estimate the very extreme quantiles using  extrapolation methods.
We set $\hat{\xi}$ to be the estimate with $\tau = 0.05$, and compute the extrapolation estimator (\ref{regression:extrapo1}) for $\tau = 0.005$, $0.001$, and $0.0001$ in Table \ref{table1-2}.
For comparison purposes, the first column reports the $\tau$-QR estimates  for $\tau = 0.005$ obtained from (\ref{regression}). This estimator cannot be calculated for the other quantile indexes considered.
We find some discrepancies between the two estimators especially for the coefficients of the negative lags at $\tau = 0.005$.
Figure \ref{fig1-4} plots the predicted values for the conditional $0.005$-quantiles in the second half of 2015 obtained from the  QR and extrapolation estimators.
The standard QR fit uses sample data that contains few observations on the extreme events, while the extrapolated fit uses the tail model and a reliably estimated conditional $0.05$-quantile coefficients to predict the magnitude of such events. The quality of this prediction clearly depends on whether the tails model is accurate.

\begin{table}[hptb]
\caption{Value-at-Risk: Extrapolation Estimators for the Quantile Regression} \label{table1-2}
\begin{tabular}{lccccc}
\hline \hline
& Regression && \multicolumn{3}{c}{Extrapolation} \\
\multicolumn{1}{c}{Variable} & estimate && \multicolumn{3}{c}{estimate} \\
\cline{2-2} \cline{4-6}
& $\tau = 0.005$ && $\tau = 0.005$ & $\tau = 0.001$ & $\tau = 0.0001$ \\
\hline
Intercept & $-0.066$ && $-0.067$ & $-0.122$ & $\hphantom{0}{-0.274}$ \\
C lag return ($+$) & $-0.646$ && $-1.530$ & $-2.888$ & $\hphantom{0}{-6.642}$ \\
C lag return ($-$) & $-2.179$ && $-4.806$ & $-9.117$ & $-21.033$ \\
DJI lag return ($+$) & $\hphantom{-}1.583$ && $\hphantom{-}0.630$ & $\hphantom{-}1.167$ & $\hphantom{-0}2.652$ \\
DJI lag return ($-$) & $\hphantom{-}3.283$ && $\hphantom{-}2.425$ & $\hphantom{-}4.194$ & $\hphantom{-0}9.085$ \\
DJUSFN lag return ($+$) & $-0.742$ && $\hphantom{-}0.558$ & $\hphantom{-}1.597$ & $\hphantom{-0}4.470$ \\
DJUSFN lag return ($-$) & $-1.004$ && $\hphantom{-}1.107$ & $\hphantom{-}2.547$ & $\hphantom{-0}6.526$ \\
\hline
\end{tabular}
\end{table}

\subsection{Application 2: Contagion of Financial Risk}

We consider an application to  contagion of financial risk between commercial banks.
The response variable $Y_t$ is the daily return of the Citigroup stock (C), and the covariates $X_{1t}$, $X_{2t}$, and $X_{3t}$ are the contemporaneous daily returns of the stocks of other banks, namely, Bank of America (BAC), JPMorgan Chase \& Co.\ (JPM), and Wells Fargo \& Co.\ (WFC).
As in the previous section, we estimate the quantiles of $Y_t$ conditional on $X_t = (1, X_{1t}^+, X_{1t}^-, X_{2t}^+, X_{2t}^-, X_{3t}^+, X_{3t}^-)$ using 1,738 daily observations covering the  period from January 1, 2009 to November 30, 2015.

\medskip

\begin{figure}
\includegraphics[width=\textwidth]{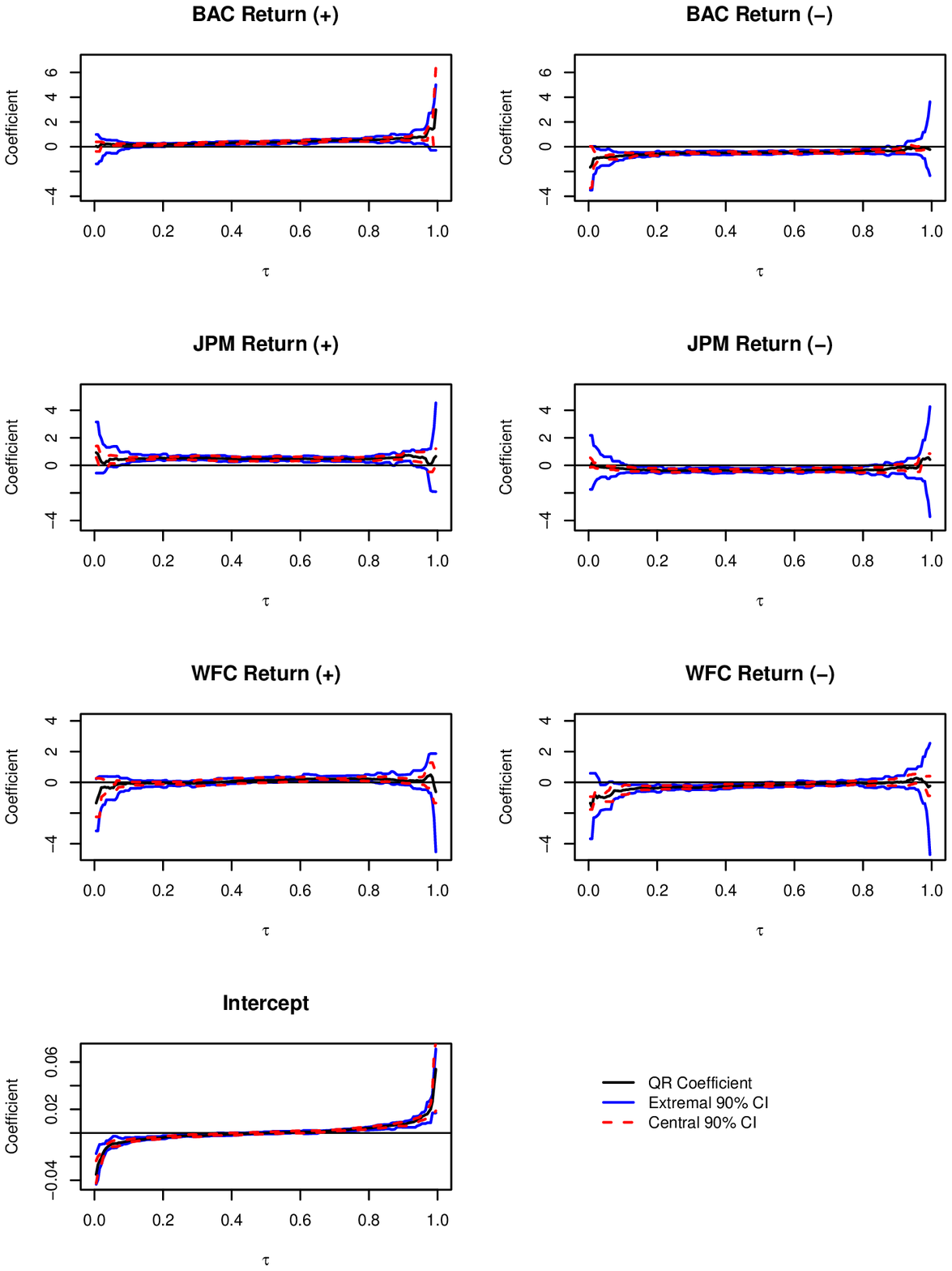}
\caption{Financial Contagion: QR coefficient estimates and 90\% pointwise CIs. The response variable is the daily Citigroup return  from January 1, 2009 to November 30, 2015. The solid lines depict extremal CIs and the dashed lines depict normal CIs.} \label{fig2-1}
\end{figure}

\begin{figure}
\includegraphics[width=\textwidth]{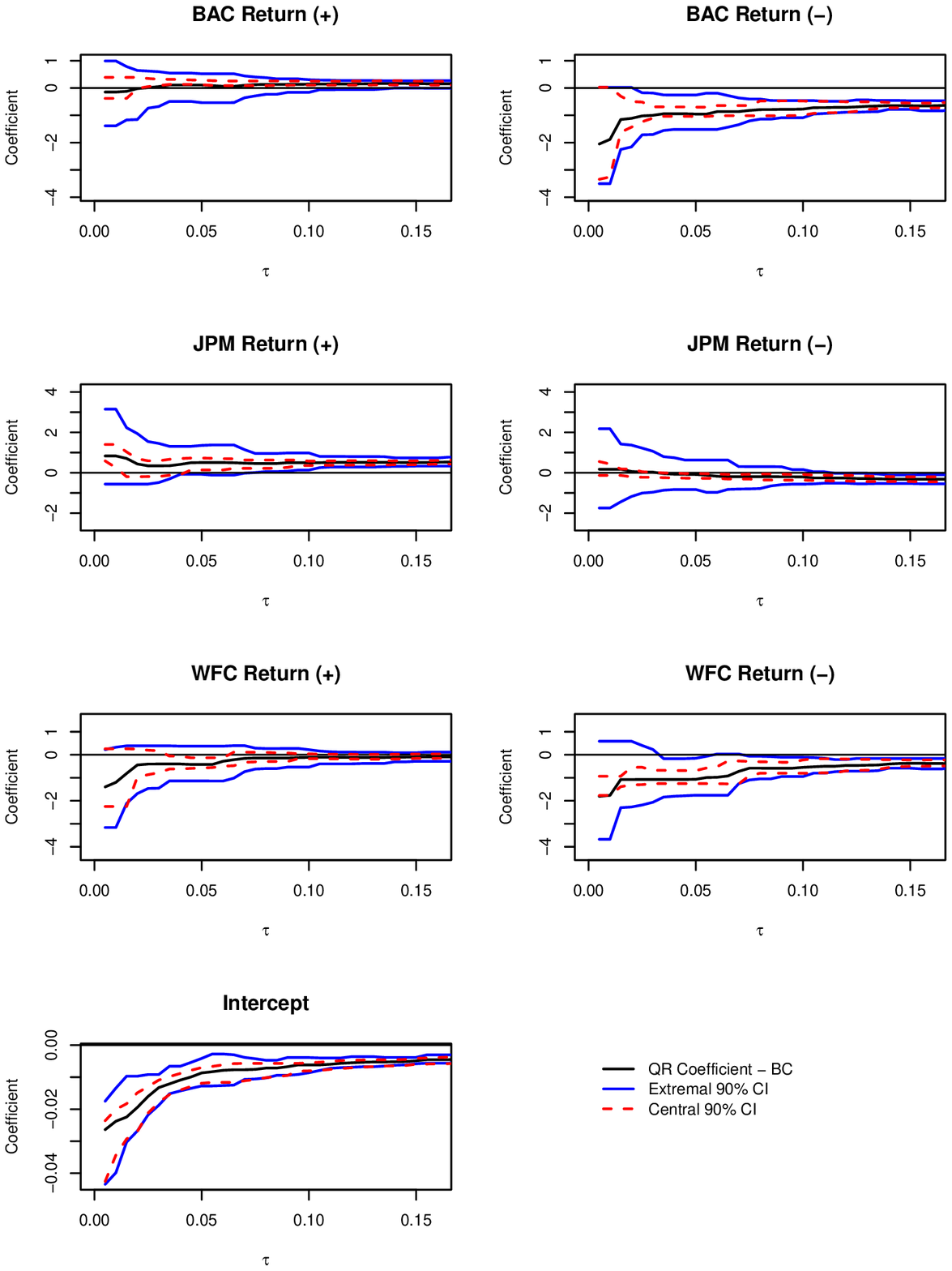}
\caption{Financial Contagion: Bias-corrected QR coefficient estimates and 90\% pointwise CIs for low quantiles. The response variable is the daily Citigroup return  from January 1, 2009 to November 30, 2015. The solid lines depict extremal CIs and the dashed lines depict normal CIs.} \label{fig2-2}
\end{figure}

Figure \ref{fig2-1} plots the QR estimates $\hat \beta(\tau)$ along with 90\% pointwise CIs. The solid lines represent the extremal CIs and the dashed lines the normal CIs. The extremal CIs are computed by the extremal subsampling method described in Algorithm \ref{alg:ces} with the subsample size $b = \lfloor 50 + \sqrt{1,738} \rfloor = 91$  and the number of simulations $S = 500$. We use the SN-QR statistic with spacing parameter $p=5$. 
The normal CIs are based on the normal approximation with the standard errors computed with the method proposed by \citet{powell:1991}.
Figure \ref{fig2-2} plots the median bias-corrected QR estimates along with 90\% pointwise CIs for the lower tail. The bias correction is also implemented using extremal subsampling with the same specifications.

\medskip

We find a significant effect of Bank of America's risk on Citigroup's risk.
Observe that the coefficient of BAC ($+$) is  positive and that of BAC ($-$) is  negative across most of the quantiles.
This tells that  BAC and C hold similar portfolios and that  there might be a direct contagion of BAC's risk to C's risk (negative return of BAC is likely to cause negative return of C). Similar observation holds for JPM's risk onto C's risk.
However, there are no such contagion effect of WFC's risk onto C's. In fig.  \ref{fig2-2} we see that the negative return of Bank of America's stock has a large effect on the extreme low quantile of C's return, while its positive return has no significant effect. This indicates that Bank of America's risk has an asymmetric and large impact on its competitor. As in the value at risk application, we find that the normal and extremal CIs are similar in the central region, while the normal CIs are narrower than the extremal CI in the tails, especially for $\tau < 0.05$.

\medskip


\begin{figure}
\includegraphics[width=\textwidth]{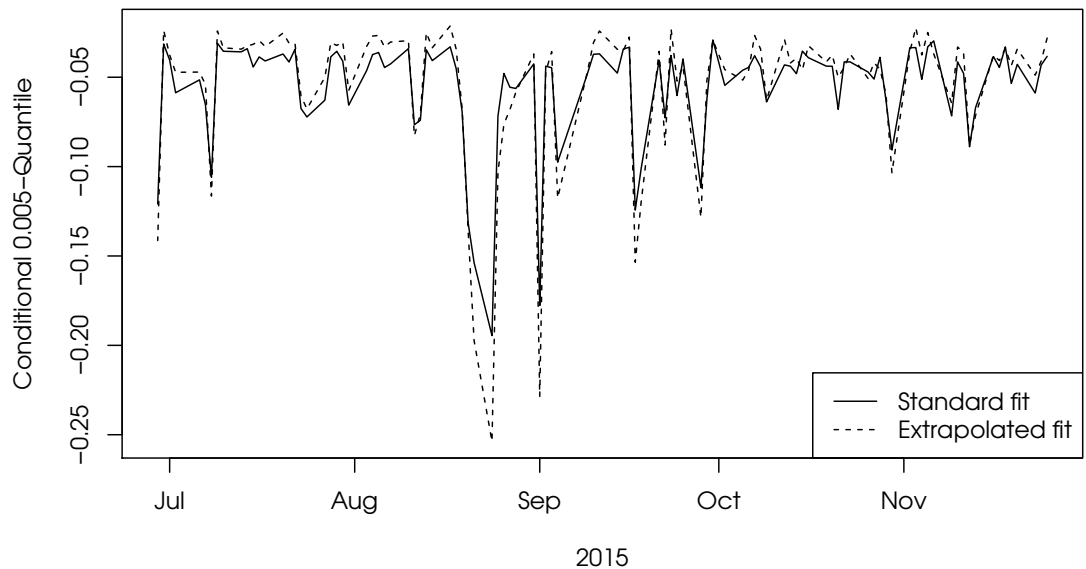}
\caption{Financial Contagion: Extrapolation vs Standard Estimates of Conditional Quantiles of the daily Citigroup return between July 1, 2015 and November 30, 2015.} \label{fig2-4}
\end{figure}

Table \ref{table2-1} reports the estimates of the EV index $\xi$ obtained by the Hill estimator (\ref{regression:hill}), together with bias corrected estimates and  90\% CIs based on (\ref{regression:hill:dist}), which were obtained using the QR extremal bootstrap of Algorithm \ref{alg:ceb} with $S=500$ applied to the Hill estimator.
Again we find estimates significantly greater than zero confirming that stock returns have thick lower tails relative to the normal distribution.  
Table \ref{table2-2} shows the estimates of the QR coefficients for very low quantiles obtained from QR and the extrapolation estimator (\ref{regression:extrapo1})  with $\hat{\xi} = 0.263,$ the estimate from table \ref{table2-1} for $\tau = 0.05$. 
The largest difference between the regression and extrapolation estimates occur for the WFC's return. Here we find a large negative coefficient for the return (-) that indicates that there might be contagion of financial risk from WFC to C at very low quantiles. 
Figure \ref{fig2-4} contrasts the predicted values for the conditional $0.005$-quantiles in the second half of 2015 obtained from the  QR and extrapolation estimators. Overall, the two methods produce similar estimates, although the extrapolated estimator predicts deeper troughs in the quantiles.

\begin{table}[hptb]
\caption{Financial Contagion: Hill Estimation Results for the EV Index} \label{table2-1}
\begin{tabular}{lccc}
\hline \hline
& Estimate & Bias-Corrected & 90\% Confidence \\
& & Estimate & Interval \\
\hline
$\tau = 0.01$ & $0.263$ & $0.255$ & $[0.086, \ 0.461]$ \\
$\tau = 0.05$ & $0.646$ & $0.611$ & $[0.468, \ 1.000]$ \\
$\tau = 0.1$ & $0.500$ & $0.357$ & $[0.276, \ 0.441]$ \\
\hline\hline
\end{tabular}
\end{table}

\begin{table}[hptb]
\caption{Financial Contagion: Extrapolation Estimators for the Quantile Regression} \label{table2-2}
\begin{tabular}{lccccc}
\hline \hline
& Regression && \multicolumn{3}{c}{Extrapolation} \\
\multicolumn{1}{c}{Variable} & estimate && \multicolumn{3}{c}{estimate} \\
\cline{2-2} \cline{4-6}
& $\tau = 0.005$ && $\tau = 0.005$ & $\tau = 0.001$ & $\tau = 0.0001$ \\
\hline
Intercept & $-0.035$ && $-0.022$ & $-0.037$ & $\hphantom{0}{-0.074}$ \\
BAC return ($+$) & $\hphantom{-}0.042$ && $\hphantom{-}0.203$ & $\hphantom{-}0.227$ & $\hphantom{-0}0.285$ \\
BAC return ($-$) & $-1.657$ && $-1.487$ & $-2.202$ & $\hphantom{0}{-3.928}$ \\
JPM return ($+$) & $\hphantom{-}0.931$ && $\hphantom{-}0.328$ & $\hphantom{-}0.205$ & $\hphantom{0}{-0.091}$ \\
JPM return ($-$) & $-0.149$ && $\hphantom{-}0.165$ & $\hphantom{-}0.542$ & $\hphantom{-0}1.451$ \\
WFC return ($+$) & $-1.350$ && $-1.775$ & $-3.437$ & $\hphantom{0}{-7.449}$ \\
WFC return ($-$) & $-1.352$ && $-3.279$ & $-5.960$ & $-12.430$ \\
\hline
\end{tabular}
\end{table}

\bibliographystyle{elsarticle-harv}
\bibliography{bibtex_handbook}

\begin{thebibliography}{52}
\expandafter\ifx\csname natexlab\endcsname\relax\def\natexlab#1{#1}\fi
\expandafter\ifx\csname url\endcsname\relax
  \def\url#1{\texttt{#1}}\fi
\expandafter\ifx\csname urlprefix\endcsname\relax\def\urlprefix{URL }\fi

\bibitem[{Abrevaya(2001)}]{abrevaya}
Abrevaya, J., 2001. The effect of demographics and maternal behavior on the
  distribution of birth outcomes. Empirical Economics 26~(1), 247--259.

\bibitem[{Aigner et~al.(1976)Aigner, Amemiya, and Poirier}]{aigner:1976}
Aigner, D.~J., Amemiya, T., Poirier, D.~J., 1976. On the estimation of
  production frontiers: maximum likelihood estimation of the parameters of a
  discontinuous density function. International Economic Review 17~(2),
  377--396.

\bibitem[{Aigner and fan Chu(1968)}]{aigner:1968}
Aigner, D.~J., fan Chu, S., 1968. On estimating the industry production
  function. American Economic Review 58, 826--839.

\bibitem[{Altonji et~al.(2012)Altonji, Ichimura, and Otsu}]{altonji:2012}
Altonji, J.~G., Ichimura, H., Otsu, T., 2012. Estimating derivatives in
  nonseparable models with limited dependent variables. Econometrica 80~(4),
  1701--1719.

\bibitem[{Arrow et~al.(1951)Arrow, Harris, and Marschak}]{arrow:ss}
Arrow, K.~J., Harris, T., Marschak, J., 1951. Optimal inventory policy.
  Econometrica 19~(3), 205--272.

\bibitem[{Bertail et~al.(2004)Bertail, Haefke, Politis, and
  White}]{bertail:2004}
Bertail, P., Haefke, C., Politis, D.~N., White, H., 2004. A subsampling
  approach to estimating the distribution of diverging extreme statistics with
  applications to assessing financial market risks. Journal of Econometrics
  120~(2), 295--326.

\bibitem[{Bickel and Freedman(1981)}]{bickel}
Bickel, P., Freedman, D., 1981. Some asymptotic theory for the bootstrap.
  Annals of Statistics 9, 1196--1217.

\bibitem[{Chernozhukov(1998)}]{chernozhukov:1998}
Chernozhukov, V., 1998. Nonparametric extreme regression quantiles, working
  paper, Standord Univ. Presented at Princeton Econometrics Seminar, December
  1998.

\bibitem[{Chernozhukov(2005)}]{victor:annals}
Chernozhukov, V., 2005. Extremal quantile regression. Ann. Statist. 33~(2),
  806--839.

\bibitem[{Chernozhukov and Du(2008)}]{chernozhukov-du:2008}
Chernozhukov, V., Du, S., 2008. extremal quantiles and value-at-risk. In:
  Durlauf, S.~N., Blume, L.~E. (Eds.), The New Palgrave Dictionary of
  Economics. Palgrave Macmillan, Basingstoke.

\bibitem[{Chernozhukov and Fern\'andez-Val(2005)}]{chernozhukov:2005}
Chernozhukov, V., Fern\'andez-Val, I., 2005. Subsampling inference on quantile
  regression processes. Indian Journal of Statistics 67, 253--276.

\bibitem[{Chernozhukov and Fern\'andez-Val(2011)}]{chernozhukov:2011}
Chernozhukov, V., Fern\'andez-Val, I., 2011. Inference for extremal conditional
  quantile models, with an application to market and birthweight risks. Review
  of Economic Studies 78, 559--589.

\bibitem[{Chernozhukov and Umantsev(2001)}]{vl}
Chernozhukov, V., Umantsev, L., 2001. Conditional value-at-risk: Aspects of
  modeling and estimation. Empirical Economics 26~(1), 271--293.

\bibitem[{de~Haan(1970)}]{dehaan:1970}
de~Haan, L., 1970. On Regular Variation and Its Applications to the Weak
  Convergence. Mathematical Centre Tract 32, Mathematical Centre, Amsterdam,
  Holland.

\bibitem[{Dekkers and de~Haan(1989)}]{Dekkers:dehaan}
Dekkers, A., de~Haan, L., 1989. On the estimation of the extreme-value index
  and large quantile estimation. Annals of Statistics 17~(4), 1795--1832.

\bibitem[{D'Haultf{\oe}uille et~al.(2015)D'Haultf{\oe}uille, Maurel, and
  Zhang}]{d'haultfoeuille:2015}
D'Haultf{\oe}uille, X., Maurel, A., Zhang, Y., 2015. Extremal quantile
  regressions for selection models and the black-white wage gap, working Paper.

\bibitem[{Donald and Paarsch(2002)}]{dp:superconsistent}
Donald, S.~G., Paarsch, H.~J., 2002. Superconsistent estimation and inference
  in structural econometric models using extreme order statistics. Journal of
  Econometrics 109~(2), 305--340.

\bibitem[{Embrechts et~al.(1997)Embrechts, Kl{\"u}ppelberg, and
  Mikosch}]{embrechts:1997}
Embrechts, P., Kl{\"u}ppelberg, C., Mikosch, T., 1997. Modelling extremal
  events 33.

\bibitem[{Engle and Manganelli(2004)}]{engle:2004}
Engle, R.~F., Manganelli, S., 2004. Cariar: Conditional autoregressive value at
  risk by regression quantiles. Journal of Business and Economic Statistics
  22~(4), 367--381.

\bibitem[{Fama(1965)}]{fama:1965}
Fama, E.~F., 1965. The behavior of stock market prices. Journal of Business 38,
  34--105.

\bibitem[{Feigin and Resnick(1994)}]{feigin:1994}
Feigin, P.~D., Resnick, S.~I., 1994. Limit distributions for linear programming
  time series estimators. Stochastic Processes and their Applications 51,
  135--165.

\bibitem[{Flinn and Heckman(1982)}]{flinn:1982}
Flinn, C.~J., Heckman, J.~J., 1982. New methods for analyzing structural models
  of labor force dynamics. Journal of Econometrics 18~(1), 115--168.

\bibitem[{Fox and Rubin(1964)}]{fox:rubin}
Fox, M., Rubin, H., 1964. Admissibility of quantile estimates of a single
  location parameter. Annals of Mathematical Statistics 35, 1019--1030.

\bibitem[{Gnedenko(1943)}]{gnedenko:1943}
Gnedenko, B., 1943. Sur la distribution limit\'e du terme d' une s\'erie
  al\'etoire. Annals of Mathematics 44, 423--453.

\bibitem[{Gutenbrunner et~al.(1993)Gutenbrunner, Jure{\v{c}}kov{\'a}, Koenker,
  and Portnoy}]{gjkp:1993}
Gutenbrunner, C., Jure{\v{c}}kov{\'a}, J., Koenker, R., Portnoy, S., 1993.
  Tests of linear hypotheses based on regression rank scores. J. Nonparametr.
  Statist. 2~(4), 307--331.
\newline\urlprefix\url{http://dx.doi.org/10.1080/10485259308832561}

\bibitem[{He et~al.(2016)He, Cheng, and Tong}]{he:2016}
He, F., Cheng, Y., Tong, T., 2016. Estimation of extreme conditional quantiles
  through an extrapolation of intermediate regression quantiles. Statistics and
  Probability Letters 113, 30--37.

\bibitem[{Hill(1975)}]{hill}
Hill, B.~M., 1975. A simple general approach to inference about the tail of a
  distribution. Annals of Statistics 3~(5), 1163--1174.

\bibitem[{Jansen and de~Vries(1991)}]{jansen:1991}
Jansen, D.~W., de~Vries, C.~G., 1991. On the frequency of large stock returns:
  Putting booms and busts into perspective. Review of Economics and Statistics
  73, 18--24.

\bibitem[{Jure{\v{c}}kov{\'a}(1999)}]{jureckova:1999}
Jure{\v{c}}kov{\'a}, J., 1999. Regression rank-scores tests against
  heavy-tailed alternatives. Bernoulli 5~(4), 659--676.
\newline\urlprefix\url{http://dx.doi.org/10.2307/3318695}

\bibitem[{Jure{\v{c}}kov{\'a}(2016)}]{jureckova:2016}
Jure{\v{c}}kov{\'a}, J., 2016. Averaged extreme regression quantile. Extremes
  19~(1), 41--49.
\newline\urlprefix\url{http://dx.doi.org/10.1007/s10687-015-0232-2}

\bibitem[{Knight(2001)}]{knight:linear}
Knight, K., 2001. Limiting distributions of linear programming estimators.
  Extremes 4~(2), 87--103.

\bibitem[{Koenker(2016)}]{koenker:2016}
Koenker, R., 2016. quantreg: Quantile Regression. R package version 5.21.
\newline\urlprefix\url{https://CRAN.R-project.org/package=quantreg}

\bibitem[{Koenker and Bassett(1978)}]{koenker:1978}
Koenker, R., Bassett, G.~S., 1978. Regression quantiles. Econometrica 46,
  33--50.

\bibitem[{Laplace(1818)}]{laplace:1818}
Laplace, P.-S., 1818. Th\'eorie analytique des probabilit\'es. \'Editions
  Jacques Gabay (1995), Paris.

\bibitem[{Leadbetter et~al.(1983)Leadbetter, Lindgren, and
  Rootz{\'e}n}]{leadbetter}
Leadbetter, M.~R., Lindgren, G., Rootz{\'e}n, H., 1983. Extremes and related
  properties of random sequences and processes. Springer-Verlag, New
  York-Berlin.

\bibitem[{Longin(1996)}]{longin:1996}
Longin, F.~M., 1996. The asymptotic distribution of extreme stock market
  returns. Journal of Business 69~(3), 383--408.

\bibitem[{Mandelbrot(1963)}]{mandelbrot:1963}
Mandelbrot, M., 1963. The variation of certain speculative prices. Journal of
  Business 36, 394--419.

\bibitem[{Meyer(1973)}]{meyer:1973}
Meyer, R.~M., 1973. A poisson-type limit theorem for mixing sequences of
  dependent ``rare" events. Annals of Probability 1, 480--483.

\bibitem[{Pickands(1975)}]{pickands:1975}
Pickands, III, J., 1975. Statistical inference using extreme order statistics.
  Annals of Statistics 3, 119--131.

\bibitem[{Politis et~al.(1999)Politis, Romano, and Wolf}]{politis:1999}
Politis, D.~N., Romano, J.~P., Wolf, M., 1999. Subsampling. New York:
  Springer-Verlag.

\bibitem[{Portnoy and Jure{\u{c}}kov{\'a}(1999)}]{portnoy:jur}
Portnoy, S., Jure{\u{c}}kov{\'a}, J., 1999. On extreme regression quantiles.
  Extremes 2~(3), 227--243.

\bibitem[{Portnoy and Koenker(1989)}]{portnoy-koenker:1989}
Portnoy, S., Koenker, R., 1989. Adaptive {$L$}-estimation for linear models.
  Ann. Statist. 17~(1), 362--381.
\newline\urlprefix\url{http://dx.doi.org/10.1214/aos/1176347022}

\bibitem[{Powell(1984)}]{powell:1984}
Powell, J.~L., 1984. Least absolute deviations estimation for the censored
  regression model. Journal of Econometrics 25, 303--325.

\bibitem[{Powell(1991)}]{powell:1991}
Powell, J.~L., 1991. Estimation of monotonic regression models under quantile
  restrictions. Nonparametric and semiparametric methods in
  Econometrics,(Cambridge University Press, New York, NY), 357--384.

\bibitem[{Praetz(1972)}]{praetz:1972}
Praetz, V., 1972. The distribution of share price changes. Journal of Business
  45~(1), 49--55.

\bibitem[{Sen(1973)}]{sen:1973}
Sen, A., 1973. On economic inequality.

\bibitem[{Smith(1994)}]{smith:1994}
Smith, R.~L., 1994. Nonregular regression. Biometrika 81~(1), 173--183.
\newline\urlprefix\url{http://dx.doi.org/10.1093/biomet/81.1.173}

\bibitem[{Timmer(1971)}]{timmer}
Timmer, C.~P., 1971. Using a probabilistic frontier production function to
  measure technical efficiency. Journal of Political Economy 79, 776--794.

\bibitem[{Wang et~al.(2012)Wang, Li, and He}]{wang:2012}
Wang, H., Li, D., He, X., 2012. Estimation of high conditional quantiles for
  heavy-tailed distributions. Journal of the American Statistical Association
  107, 1453--1464.

\bibitem[{White(2001)}]{white:2001}
White, H., 2001. Asymptotic Theory for Econometricians, revised Edition. New
  York: Academic Press.

\bibitem[{Zhang(2015)}]{zhang:2015}
Zhang, Y., 2015. Extremal quantile treatment effects, job Market Paper.

\bibitem[{Zipf(1949)}]{zipf:1949}
Zipf, G., 1949. Human Behavior and the Principle of Last Effort. Cambridge, MA:
  Addison-Wesley.

\end{thebibliography}

\end{document}